\documentclass[12pt,draftclsnofoot,onecolumn]{IEEEtran}
\usepackage{amssymb,graphicx,amsmath,psfig,cite,float,xcolor,setspace}

\newcommand{\bi}{\begin{itemize}}
\newcommand{\ei}{\end{itemize}}

\newcommand{\be}{\begin{enumerate}}
\newcommand{\ee}{\end{enumerate}}
\newcommand{\bd}{\begin{description}}
\newcommand{\ed}{\end{description}}
\newcommand{\bc}{\begin{center}}
\newcommand{\ec}{\end{center}}
\newcommand{\bt}{\begin{tabbing}}
\newcommand{\et}{\end{tabbing}}
\newcommand{\bfig}{\begin{figure}}
\newcommand{\efig}{\end{figure}}
\newcommand{\beq}{\begin{equation}}
\newcommand{\beqarr}{\begin{eqnarray}}
\newcommand{\beqarrn}{\begin{eqnarray*}}
\newcommand{\eeq}{\end{equation}}
\newcommand{\eeqarr}{\end{eqnarray}}
\newcommand{\eeqarrn}{\end{eqnarray*}}
\newcommand{\bflr}{\begin{flushright}\vspace{-0.2in}}
\newcommand{\eflr}{\end{flushright}}
\newcommand{\bsub}{\begin{subequations}}
\newcommand{\esub}{\end{subequations}}
\newcommand{\barr}{\begin{array}}
\newcommand{\earr}{\end{array}}
\newcommand{\nn}{\nonumber}

\def\undb#1{\mbox{\bf{#1}}}

\def\dn{\stackrel{\scriptscriptstyle \triangle}{=}}

\def\arg{\mbox{arg}}

\begin{document}
\title{
RIS-Aided Index Modulation with
Greedy Detection over Rician Fading Channels
}
%\author{Author 1
%        and~Author 2
\author{Aritra Basu, Soumya~P.~Dash,~\IEEEmembership{Member,~IEEE}, Aryan Kaushik,~\IEEEmembership{Member,~IEEE}, \\ Debasish Ghose,~\IEEEmembership{Senior Member,~IEEE}, Marco Di Renzo,~\IEEEmembership{Fellow,~IEEE}, and \\ Yonina C. Eldar,~\IEEEmembership{Fellow,~IEEE}

\thanks{A. Basu and S. P. Dash are with the School of Electrical Sciences, Indian
Institute of Technology Bhubaneswar, Argul, Khordha, 752050
India e-mail: (21ec06008@iitbbs.ac.in, soumyapdashiitbbs@gmail.com).}
\thanks{A. Kaushik is with the School of Engineering and Informatics, University of Sussex, Brighton, UK, e-mail: aryan.kaushik@sussex.ac.uk.}
\thanks{D. Ghose is with the School of Economics, Innovation, and Technology, Kristiania University College, Bergen, Norway 5022, e-mail: debasish.ghose@kristiania.no.}
\thanks{M. D. Renzo is with the Laboratoire des Signaux et Systmes, CNRS, CentraleSuplec, University Paris Sud, Universit Paris-Saclay, 91192 Paris, France, e-mail: marco.direnzo@centralesupelec.fr.}
\thanks{Y. C. Eldar is with Weizmann Institute of Science, Rehovot 7610001,
Israel, e-mail: yonina.eldar@weizmann.ac.il.}
\vspace{-1.5cm}
}
\maketitle
\doublespacing
%%%%%%%%%%%%%%%%%%%%%%%%%%%%%%%%%%%%%%%%%%%%%%%%%%%%%%%%%%%%%%%%%%%%%%%%%%%%%%%%%%%%%%%%%%%
%%%%%%%%%%%%%%%%%%%%%%%%%%%%%%%%%%%%%%%%%%%%%%%%%%%%%%%%%%%%%%%%%%%%%%%%%%%%%%%%%%%%%%%%%%%
\begin{abstract}
Index modulation schemes for reconfigurable intelligent surfaces (RIS)-assisted systems are envisioned as promising technologies for fifth-generation-advanced and sixth-generation (6G) wireless communication systems to enhance various system capabilities such as coverage area and network capacity. In this paper, we consider a receive diversity RIS-assisted wireless communication system employing IM schemes, namely, space-shift keying (SSK) for binary modulation and spatial modulation (SM) for $M$-ary modulation for data transmission. The RIS lies in close proximity to the transmitter, and the transmitted data is subjected to a fading environment with a prominent line-of-sight component modeled by a Rician distribution. A receiver structure based on a greedy detection rule is employed to select the receive diversity branch with the highest received signal energy for demodulation. The performance of the considered system is evaluated by obtaining a series-form expression for the probability of erroneous index detection (PED) of the considered target antenna using a characteristic function approach. In addition, closed-form and asymptotic expressions at high and low signal-to-noise ratios (SNRs) for the bit error rate (BER) for the SSK-based system, and the SM-based system employing $M$-ary phase-shift keying and $M$-ary quadrature amplitude modulation schemes, are derived. The dependencies of the system performance on the various parameters are corroborated via numerical results. The asymptotic expressions and results of PED and BER at high and low SNR values lead to the observation of a performance saturation and the presence of an SNR value as a point of inflection, which is attributed to the greedy detector's structure.
\end{abstract}

%%%%%%%%%%%%%%%%%%%%%%%%%%%%%%%%%%%%%%%%%%%%%%%%%%%%%%%%%%%%%%%%%%%%%%%%%%%%%%%%%%%%%%%%%%%
%%%%%%%%%%%%%%%%%%%%%%%%%%%%%%%%%%%%%%%%%%%%%%%%%%%%%%%%%%%%%%%%%%%%%%%%%%%%%%%%%%%%%%%%%%%
%%%%%%%%%%%%%%%%%%%%%%%%%%%%%%%%%%%%%%%%%%%%%%%%%%%%%%%%%%%%%%%%%%%%%%%%%%%%%%%%%%%%%%%%%%%
%%%%%%%%%%%%%%%%%%%%%%%%%%%%%%%%%%%%%%%%%%%%%%%%%%%%%%%%%%%%%%%%%%%%%%%%%%%%%%%%%%%%%%%%%%%
\section{Introduction}
Fifth-generation (5G) wireless communication systems promise to achieve high spectrum efficiency and energy efficiency that has led to a new vision of mobile communications, broadly catering to use cases with requirements of enhanced mobile broadband, ultra-reliability, and low-latency for massive machine-type communications \cite{psu}. Since it becomes very challenging for a single enabling technology to cater to all these demands, researchers, in a bid to make the communication systems future ready, have started to explore advanced 5G and 6th-generation (6G) envisioned technologies.
New user requirements, applications, use cases, and networking trends are expected to thrive, which will necessitate new communication paradigm shifts, especially at the physical layer \cite{hgr22, iggy44}. Within this context for the development of next-generation wireless communication technologies \cite{heq11, sav11, aaq1}, there has been increasing interest in controlling the properties of the physical channel/medium, which has led to the popularity of reconfigurable intelligent surfaces (RIS) \cite{falturef, dixcy29, sv22}. Furthermore, the application of other new technologies, such as index modulation (IM) to RIS-assisted systems, has piqued the interest of researchers over recent years.
%%%%%%%%%%%%%%%%%%%%%%%%%%%%%%%%%%%%%%%%%%%%%%%%%%%%%%%%%%%%%%%%%%%%%%%%%%%%%%%%%%%%%%%%%%%
%%%%%%%%%%%%%%%%%%%%%%%%%%%%%%%%%%%%%%%%%%%%%%%%%%%%%%%%%%%%%%%%%%%%%%%%%%%%%%%%%%%%%%%%%%%
\subsection{Literature Review}
RIS, consisting of reconfigurable meta-surfaces, influence the wireless medium/channel in which they are placed. The ease of deployment of thin artificial films of RIS results in a reduction in implementation cost and system complexity \cite{lopp234, res29, pops22}. Moreover, RIS technology, considered a passive form of a relay \cite{vge22, asw:21, llop22, llop21, MaBhPa:14}, does not require a dedicated energy source to cater to different RF processing, encoding, decoding, or re-transmission, and are thus, efficient over existing multiple-input multiple-output (MIMO), beamforming, amplify-and-forward relaying, and backscatter communication paradigms \cite{vetu2910}. Generally, RIS-assisted systems tend to perform better with fewer antennas than existing MIMO systems, thus significantly reducing implementation costs \cite{dixcy, wea61}. Furthermore, the reflection characteristics of these surfaces may be controlled by software; hence a RIS is also referred to as a software-defined surface and can be employed as a substitute for traditional beamforming techniques \cite{dswqrsd}.

Another vital aspect of beyond 5G systems is the use of IM \cite{ffeq22, huu21}, which specifies a type of modulation technique relying on some form of activation states for information embedding, carried out in different domains like space, time, and frequency slots, or even a combination of them. The basic principle of IM is to separate the information bits into index and constellation bits. The former specifies the portion of the active radio resources (antennas and sub-carriers), and the latter bits are used for mapping conventional constellation symbols to be carried by the active resources \cite{turkiye29, dixcydoctor}. In this category, space-shift keying (SSK) and spatial modulation (SM) exploit the spatial-constellation diagram for data modulation, resulting in a low-complexity modulation for multiple-antenna systems, and they outperform conventional modulation schemes in terms of spectral efficiency \cite{uaiguv:45, qw:33, henry, puri}. Though the basic principle behind index-modulated SSK or SM schemes is similar in terms of data modulation, SM-based index-modulated systems are more complex due to the additional requirement of a modulation system/scheme in the form of amplitude/phase modulation.

Several studies have been proposed in the literature for these potential next-generation communication technologies for RIS-assisted IM scheme-based system models. The authors in \cite{vetu2910} proposed three IM-based communication system architectures: IM for the source transmit antennas; IM for the destination receive antennas; and IM for the RIS region proposing reflection modulation-based beam patterns, and have analyzed the system in terms of its outage probability. A hybrid concept employing IM and metasurface modulation is studied in \cite{res29} to improve the system's performance. An RIS grouping-based IM to enhance spectral efficiency and improve bit error rate is presented in \cite{gcct21}. The authors in \cite{zxc1} introduced the concept of reflection modulation, whereas the authors in \cite{aq245} proposed a non-coherent receiver for a reflection modulation-based RIS-assisted system to further reduce the hardware complexity of the wireless communication system. The performance loss associated with a RIS-assisted SSK modulated system with a blind receiver is examined in \cite{wb:223}. The authors in \cite{vetu2910} and \cite{MaBhPa:14} have considered SSK and SM schemes for a receive diversity RIS-assisted communication system wherein they proposed a greedy detector for detecting the diversity branch with maximum received energy and utilized the same for data demodulation. With this system model, \cite{vetu2910} obtained upper bounds on the system performance, and the authors in \cite{MaBhPa:14} derived closed-form expressions for the same performance metrics.

\subsection{Contributions}
RIS systems are generally utilized owing to the lack of a line of sight (LoS) communication path between the transmitter and the receiver end, thus resulting in an LoS path being present between the transmitter-RIS network and the RIS-receiver network. However, most previously carried out works for IM-based RIS-assisted systems do not consider this scenario where the channel gains generally follow Rician fading \cite{jfk2214, falturef2, avqq2308}, thus failing to model communication scenarios that possess significant or dominant LoS components.
We consider an SSK/SM-based RIS-assisted wireless system in which the envelopes of the channel gains involved follow Rician distributions, thus being able to properly model the scenario between the RIS and the receiver that possesses a dominant LoS path. The SSK system is considered for binary modulation schemes, and the SM system is considered for the transmission of $M$-ary modulated data symbols. Further motivated by the studies in \cite{vetu2910, MaBhPa:14}, we employ a greedy detector at the receiver and study the systems' performance in terms of the probability of erroneous detection (PED) of the corresponding antenna of interest.
The contributions of the work are summarized as follows:
\begin{itemize}
\item A receive diversity wireless communication system subject to Rician fading channels is considered wherein the transmitter, lying in close proximity to the RIS, employs an IM modulation scheme based on SSK or $M$-ary phase-shift keying ($M$-PSK)/$M$-ary quadrature amplitude modulation (QAM) constellation-based SM scheme.
\item A greedy detector structure \cite{MaBhPa:14} is employed for the considered RIS-assisted wireless system, which relies on the maximum energy of the received signals at the receive diversity branches to be selected for demodulation without the need for channel estimation.
\item An analytical framework based on a characteristic function (c.f.) approach is proposed to obtain exact closed-form expressions for the PED of the target receive diversity antenna for the RIS-assisted SSK and SM systems.
\item Expressions for the symbol error probability (SEP) of both systems are derived, based on which the asymptotic expressions at high and low signal-to-noise ratios (SNR) provide insight into the dependency of the performance of the system on various system parameters.
\item Numerical results are presented to study the effect of the system parameters on the PED and the bit error rate (BER) performance, which showcase the dependency of the LoS component on these metrics.
\end{itemize}

The rest of the paper is organized as follows. The model of the SSK-based and SM-based RIS-assisted systems and the greedy detector are presented in Section II. The analytical frameworks to derive closed-form and asymptotic expressions of the PED of the target receive diversity antenna for the considered RIS-assisted IM-based systems are given in Sections III and IV. The corresponding derivation to obtain the exact and asymptotic expressions of the SEP of the system is provided in Section V. Section VI presents numerical results corroborating the analytical studies, followed by concluding remarks provided in Section VII.
%%%%%%%%%%%%%%%%%%%%%%%%%%%%%%%%%%%%%%%%%%%%%%%%%%%%%%%%%%%%%%%%%%%%%%%%%%%%%%%%%%%%%%%%%%%
%%%%%%%%%%%%%%%%%%%%%%%%%%%%%%%%%%%%%%%%%%%%%%%%%%%%%%%%%%%%%%%%%%%%%%%%%%%%%%%%%%%%%%%%%%%
%%%%%%%%%%%%%%%%%%%%%%%%%%%%%%%%%%%%%%%%%%%%%%%%%%%%%%%%%%%%%%%%%%%%%%%%%%%%%%%%%%%%%%%%%%%
%%%%%%%%%%%%%%%%%%%%%%%%%%%%%%%%%%%%%%%%%%%%%%%%%%%%%%%%%%%%%%%%%%%%%%%%%%%%%%%%%%%%%%%%%%%
\section{System Model}
An RIS-assisted wireless communication system is considered where the RIS consists of $N$ reconfigurable meta-surfaces and is placed in close proximity to the transmitter, similar to the concept of a RIS-AP in which the RIS can be considered as a part of the transmitter \cite{vetu2910}. The receiver, lying in the far field of the RIS to model conventional communication scenarios, comprises of $N_{R_\text{X}}$ diversity branches which obtain the signal from the transmitter via the reflections from the RIS elements. The wireless fading channel between the $i$-th reconfigurable element of the RIS and the $w$-th receive diversity branch is modeled as a complex multiplicative gain coefficient and is denoted by $h_{w,i} = \beta_{w,i} \exp \left\{- \jmath \psi_{w,i} \right\}$, where $\jmath = \sqrt{-1}$, $w=1, \ldots, N_{R_\text{X}}$, and $ i= 1, \ldots ,N$. The transmitter employs (i) SSK or (ii) $M$-ary QAM/$M$-PSK constellation-based SM scheme for index modulation, and the receiver employs a greedy detector to select the receive diversity branch for optimal performance. Furthermore, it is assumed that all the channels between the RIS and the receiver are statistically independent and identically distributed (i.i.d.) and that the transmitter has perfect knowledge of the information of the channel states. We consider the envelopes of the i.i.d. channel gains to follow Rician distributions, implying that each complex channel gain $h_{w,i}$ follows a non-zero mean complex Gaussian distribution, i.e., $h_{w,i} \sim {\mathcal{CN}} \left( \mu, \sigma_{h}^{2} \right)$, where $\undb{E} \left[ h_{w,i}\right] = \mu$ ($\undb{E}\left[ \cdot \right]$ denotes the expectation operator), $\undb{E} \left[ \left( h_{w,i} - \mu \right)^2 \right] = \sigma_h^2$, and the Rician factor of the system is denoted as $k=|\mu|^2/\sigma_{h}^{2}$.
%%%%%%%%%%%%%%%%%%%%%%%%%%%%%%%%%%%%%%%%%%%%%%%%%%%%%%%%%%%%%%%%%%%%%%%%%%%%%%%%%%%%%%%%%%%%
%%%%%%%%%%%%%%%%%%%%%%%%%%%%%%%%%%%%%%%%%%%%%%%%%%%%%%%%%%%%%%%%%%%%%%%%%%%%%%%%%%%%%%%%%%%%
\subsection{RIS-Assisted SSK System}
\begin{figure}[t]
    \centering
    \includegraphics[height=1.8in,width=4.5in]{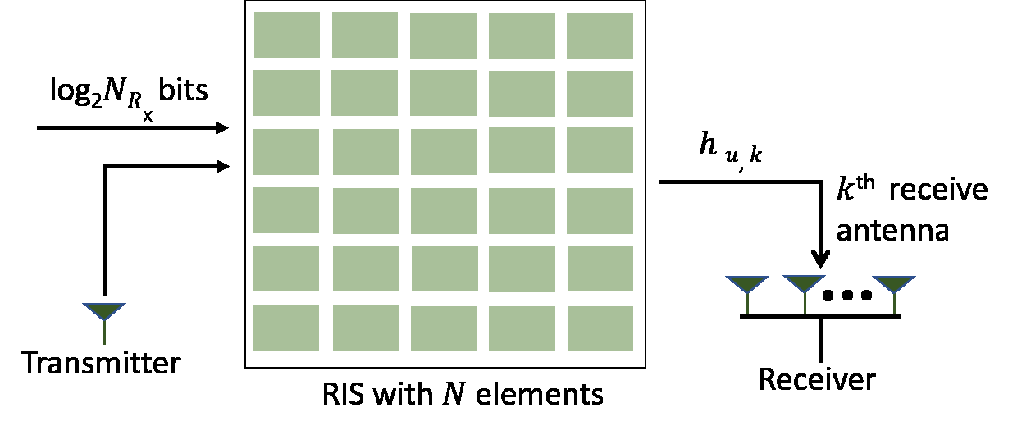}
    \caption{System model for the implementation of the RIS-assisted SSK scheme.}
    \vspace{-0.6cm}
    \label{f1}
\end{figure}
The RIS-SSK system, as shown in Fig. \ref{f1}, aims at maximizing the instantaneous received SNR at a target receive diversity branch by intelligently adjusting the phase shifts of the meta-surfaces with respect to the fading gain phases before reflecting the transmitted signal towards the receiver. The receiver performs the elementary task of determining the index of the target diversity branch by utilizing the received $\log_2 N_{R_\text{X}}$ bits.

Considering the energy of the transmitted symbol to be $\sqrt{E_s}$, the received signal at the $w$-th diversity branch can be expressed as
\beq
z_w = \sqrt{E_s} \left[ \sum_{i=1}^{N} h_{w,i} \exp \left\{\jmath \phi_i \right\} \right] + n_w \, , \, w = 1,\ldots, N_{R_\text{X}} \, ,
\label{eq1}
\eeq
where $\phi_i$ denotes the phase-shift introduced by the $i$-th meta-surface of the RIS and $n_w$ is the additive noise at the $w$-th receive diversity branch, which follows a zero-mean complex Gaussian distribution, implying that $n_w \sim {\mathcal{CN}} \left(0, N_0 \right)$ with $\undb{E} \left[ \lvert n_w \rvert^2 \right] = N_0$. Furthermore, the noise $n_w$ is statistically independent of $h_{w,i}$s and $\phi_{i} \in i = \left\{1,\ldots,N\right\}$.

From (\ref{eq1}), the instantaneous received SNR at the $w$-th receive diversity branch is given as
\beq
\gamma_w = \frac{\left|\sum\limits_{i=1}^{N}
\beta_{w,i} \exp \left\{ \jmath \left( \phi_i -\psi_{w,i} \right) \right\} \right|^2E_s}{N_0} \, , \, w = 1,\ldots, N_{R_\text{X}} \, , 
\label{eq2}
\eeq
which can be maximized by adjusting the phases of the RIS elements as $\phi_w = \psi_{w,i}$. This results in the expression of the maximum SNR at the selected $w$-th receive diversity branch, given by
\beq
\gamma_{w,max} = \frac{\left|\sum\limits_{i=1}^{N} 
\beta_{w,i}\right|^2E_s}{N_0} \, , \, w = 1,\ldots, N_{R_\text{X}} \, .
\label{eq3}
\eeq
%%%%%%%%%%%%%%%%%%%%%%%%%%%%%%%%%%%%%%%%%%%%%%%%%%%%%%%%%%%%%%%%%%%%%%%%%%%%%%%%%%%%%%%%%%%%
%%%%%%%%%%%%%%%%%%%%%%%%%%%%%%%%%%%%%%%%%%%%%%%%%%%%%%%%%%%%%%%%%%%%%%%%%%%%%%%%%%%%%%%%%%%%
\subsection{RIS-Assisted SM System}
\begin{figure}
    \centering
    \includegraphics[height=1.8in,width=4.7in]{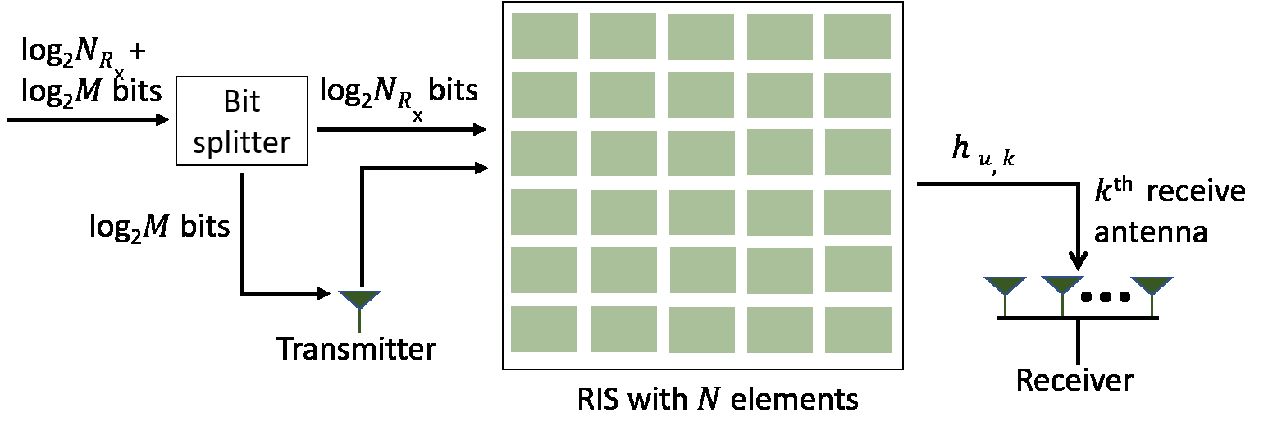}
    \caption{System model for the implementation of the RIS-assisted SM schemes.}
    \vspace{-0.6cm}
    \label{f2}
\end{figure}
The RIS-SM system, shown in Fig. \ref{f2}, aims to improve the system's spectral efficiency and maximize the instantaneous SNR at the target receive diversity branch. Thus, apart from the selection of the phase shifts of the RIS elements, the transmitter employs an $M$-ary QAM/$M$-PSK modulation scheme for data communication. Thus, as in the case of the RIS-assisted SSK system, the first set of $\log_2{N_{R_\text{X}}}$ bits is used for the determination of the index of the target antenna as well as to adjust the phase terms to maximize the instantaneous received SNR at the target antenna index and the second set of $\log_2{M}$ bits are utilized to generate the $M$-ary modulated symbols via an RF source. The received signal at the $w$-th diversity branch is then expressed as
\beq
z_w = \left[ \sum_{i=1}^{N} h_{w,i} 
\exp \left\{ \jmath \phi_i \right\} \right] v + n_w \, , \, w = 1,\ldots, N_{R_\text{X}} \, ,
\label{eq4}
\eeq
where $v$ is the data symbol belonging to the $M$-ary QAM/$M$-PSK constellation, $\undb{E} \left[|v|^2\right] = E_s$, and $n_w$ is the additive noise with $n_w \sim {\mathcal{CN}} \left(0, N_0 \right)$.
%%%%%%%%%%%%%%%%%%%%%%%%%%%%%%%%%%%%%%%%%%%%%%%%%%%%%%%%%%%%%%%%%%%%%%%%%%%%%%%%%%%%%%%%%%%%
%%%%%%%%%%%%%%%%%%%%%%%%%%%%%%%%%%%%%%%%%%%%%%%%%%%%%%%%%%%%%%%%%%%%%%%%%%%%%%%%%%%%%%%%%%%%
\subsection{Greedy Detector}
The receivers for both the cases of the RIS-SSK and the RIS-SM systems employ a greedy detector structure \cite{vetu2910, MaBhPa:14} to obtain the index of the receive diversity branch associated with the highest instantaneous received energy without the need for the knowledge of the CSI at the receiver end. Thus, the index `$w$' of the target antenna for this non-coherent receiver is selected by using the decision rule, given as
\beq
\hat{w} = \arg \max_w \left| z_w \right|^2 \, .
\label{eq5}
\eeq
Thus, the greedy detector chooses the {\em target antenna} according to the maximum instantaneous SNR at the diversity branches. In the following sections, we derive expressions for the PED of the target antenna of the RIS-assisted SSK and SM systems utilizing the greedy detector for the wireless communication system with a prominent LoS channel.
%%%%%%%%%%%%%%%%%%%%%%%%%%%%%%%%%%%%%%%%%%%%%%%%%%%%%%%%%%%%%%%%%%%%%%%%%%%%%%%%%%%%%%%%%%%%
%%%%%%%%%%%%%%%%%%%%%%%%%%%%%%%%%%%%%%%%%%%%%%%%%%%%%%%%%%%%%%%%%%%%%%%%%%%%%%%%%%%%%%%%%%%%
%%%%%%%%%%%%%%%%%%%%%%%%%%%%%%%%%%%%%%%%%%%%%%%%%%%%%%%%%%%%%%%%%%%%%%%%%%%%%%%%%%%%%%%%%%%%
%%%%%%%%%%%%%%%%%%%%%%%%%%%%%%%%%%%%%%%%%%%%%%%%%%%%%%%%%%%%%%%%%%%%%%%%%%%%%%%%%%%%%%%%%%%%
\section{PED Analysis of the RIS-Assisted SSK System}
In this section, we derive a series-form expression for the probability of erroneous receive antenna index detection of the RIS-SSK scheme using the greedy detector.
%%%%%%%%%%%%%%%%%%%%%%%%%%%%%%%%%%%%%%%%%%%%%%%%%%%%%%%%%%%%%%%%%%%%%%%%%%%%%%%%%%%%%%%%%%%%
%%%%%%%%%%%%%%%%%%%%%%%%%%%%%%%%%%%%%%%%%%%%%%%%%%%%%%%%%%%%%%%%%%%%%%%%%%%%%%%%%%%%%%%%%%%%
\vspace{-0.3cm}
\subsection{Pariwise PED Analysis}
Let $w$ and $\hat{w}$ be the indices of the target and some other non-target antenna, where the term {\em target antenna} is used for the receive diversity branch leading to the maximum received energy. Using (\ref{eq1}), the expression for the pairwise PED (PPED) of the target antenna is computed as
\beqarr
\text{Pr} \left\{ \left| z_{w_{ssk}} \right|^2 < \left| z_{\hat{w}_{ssk}} \right|^2\right\}
\! \! \! &=& \! \! \! \text{Pr} 
\left\{ \left| \sqrt{E_s} \sum_{i=1}^{N} h_{w,i} e^{\jmath \phi_i} + n_w \right| 
< \left| \sqrt{E_s} \sum_{i=1}^{N} h_{\hat{w},i} e^{\jmath \phi_i} + n_{\hat{w}} \right| \right\} \, .
\label{eq6}
\eeqarr
We wish to maximize the signal energy for the target antenna, which can be expressed as an optimization problem by substituting (\ref{eq1}) in (\ref{eq6}) as
\beqarr
\min_{\left\{\phi_i\right\}_{i=1}^N} \! \! \! \! && \! \! \! \!
\text{Pr} \left\{ \left| \sum_{i=1}^{N} \beta_{\hat{w},i}
\exp \left\{\jmath \left( \phi_i - \psi_{\hat{w_i},i} \right) \right\} \right|^2 
> \left| \sum_{i=1}^{N} \beta_{w,i}
\exp \left\{\jmath \left( \phi_i-\psi_{w_i,i} \right) \right\} \right|^2 \right\} . 
\label{eq7}
\eeqarr
It can be observed that (\ref{eq7}) is minimized by choosing the values of the phase shifts of the reflecting elements constituting the RIS as $\phi_i = \psi_{w,i}$ for $i=1, \ldots ,N$, resulting in the optimization problem simplifying to
\beq
\min_{\left\{\phi_w\right\}_{i=1}^N} \! \!
\text{Pr} \left\{ \left| \sum_{i=1}^{N} \beta_{\hat{w},i}
\exp \left\{ \jmath \left( \phi_i-\psi_{\hat{w_i},i} \right) \right\} \right|^2
> \left| \sum_{i=1}^{N} \beta_{w,i} \right|^2 \right\} ,
\label{eq8}
\eeq
resulting in the corresponding PPED being given by
\beqarr
\text{Pr} \left\{ \left| z_{w_{ssk}} \right|^2 \! < \! \left| z_{\hat{w}_{ssk}} \right|^2 \right\} 
= \text{Pr} \left\{ \left| \sqrt{E_s} \sum_{i=1}^{N} \beta_{w,i} + n_w \right|^2
\! \! < \left| \sqrt{E_s} \sum_{i=1}^{N} \beta_{\hat{w},i}
e^{ \jmath \left( \psi_{w,i} - \psi_{\hat{w},i} \right)}
+ n_{\hat{w}} \right|^2
\right\} .
\label{eq9}
\eeqarr

From the statistics of the complex channel gains considered in the model of the communication system, $\beta_{w,1},\ldots,\beta_{w,N}$ and $\beta_{\hat{w},1},\ldots,\beta_{\hat{w},N}$ are i.i.d. random variables, each following a Rician distribution. Moreover $\psi_{w,1},\ldots,\psi_{w,N}$ and $\psi_{\hat{w},1},\ldots,\psi_{\hat{w},N}$ are also i.i.d. random variables. Thus, utilizing the central limit theorem, the statistics of the expressions in (\ref{eq9}) are
\vspace{-0.2cm}
\bsub
\beq
\sqrt{E_s} \sum_{i=1}^N \beta_{\hat{w},i}
\exp \left\{ \jmath \left( \psi_{w,i}-\psi_{\hat{w},i} \right) \right\} + n_{\hat{w}}
\sim {\mathcal{CN}} \left( N \sqrt{E_s} \mu \, , \, N E_{s} \sigma_{h}^2 + N_0 \right) ,
\eeq
\vspace{-0.5cm}
\beqarr
&& \! \! \! \! \! \! \! \! \! \! \! \! \! \! \! \!
\! \! \! \! \! \! \! \! \! \! \! \! \! \! \! \!
\Re \left\{ \sqrt{E_s} \sum_{i=1}^{N} \beta_{w,i} + n_w \right\} \nn \\
&& \! \! \! \! \! \! \! \! \! \! \! \!
\sim {\mathcal{N}} 
\left( \frac{N\sigma_{h} \sqrt{\pi E_s} L_{1/2} \left(-k \right)}{2} ,
 N E_s \sigma_{h}^2
\left( 1 + k - \frac{\pi}{4} L^2_{1/2} \left(-k \right) \right) + \frac{N_0}{2} \right) ,
\label{eq10b}
\eeqarr
\label{eq10}
\esub
where $L_{1/2} (\cdot)$ denotes the Laguerre polynomial function, and
\beq
\Im \left\{ \sqrt{E_s} \sum_{i=1}^{N} \beta_{w,i} + n_{w} \right\}
\sim {\mathcal{N}} \left( 0, \frac{N_0}{2} \right) \, ,
\label{eq11}
\eeq
where $\Re \left\{ \cdot \right\}$ and $\Im\left\{ \cdot \right\}$ denote the real part and imaginary part operators, respectively.

We can alternatively express the PED in (\ref{eq9}) as
\beq
\text{Pr} \left\{ \left| z_w \right|^2 < \left| z_{\hat{w}} \right|^2 \right\}
= \Pr \left\{ X_{ssk}<Y_{ssk} \right\} \, ,
\label{eq12}
\eeq
where
$$ X_{ssk} = \left| \sqrt{E_s} \sum_{i=1}^{N} \beta_{w,i} + n_w \right|^2 \ , \
Y_{ssk} = \left| \sqrt{E_s} \sum_{i=1}^{N} \beta_{\hat{w},i}
\exp \left\{ \jmath \left( \psi_{w,i} - \psi_{\hat{w},i} \right) \right\}
+ n_{\hat{w}} \right|^2. $$
From the statistics provided in (\ref{eq10b}) and (\ref{eq11}), the variable $X$ can further be modeled as
\beq
X_{ssk} = \left| \left( W_0 + W_1 \right) + \jmath W_2 \right|^2 \, ,
\label{eq13}
\eeq
where $W_0$, $W_1$, and $W_2$ are independent random variables and the statistics of $W_0$, $W_1$, and $W_2$ are given by 
\beq
W_0 \sim {\mathcal{N}} \left(\mu_X, b \right) \ , \
 W_1,W_2 \sim {\mathcal{N}} \left(0,c \right) \, ,
\label{eq14}
\eeq
where
\beqarr
\mu_X = \frac{N\sigma_{h}\sqrt{\pi E_s}}{2}L_{1/2}(-k) \quad , \quad
b = N E_s \sigma_{h}^2 \left(1 + k - \frac{\pi}{4} L^2_{1/2}(-k) \right) \quad , \quad 
c = \frac{N_0}{2} ,
\label{eq15}
\eeqarr
using which the c.f. of $X_{ssk}$ can be expressed as
\beq
\Psi_{X_{ssk}} \left( \jmath \omega \right) 
= \frac{\exp \left\{\frac{\jmath \omega \mu_X^2}{1-2\jmath \omega \left(b+c\right)}\right\}}{\left(1-2\jmath \omega \left(b+c\right)\right)^\frac{1}{2}
\left(1-2\jmath \omega c\right)^\frac{1}{2}} \, .
\label{eq16}
\eeq
The statistics thus obtained for $X_{ssk}$ and $Y_{ssk}$ are utilized to derive the expression of the PPED of the system given by the following theorem.

{\em Theorem 1}: The expression of the PED in (\ref{eq12}) can be computed as
\beqarr
\text{Pr} \left\{X_{ssk} < Y_{ssk}\right\}
\! \! \! \! &=& \! \! \! \! 1 - \exp \left\{-\frac{k N^2 \Gamma_{av}}{N \Gamma_{av} + 1} \right\}
\sum_{\ell=0}^{\infty} \sum_{p=0}^{\infty}
\frac{\left(-1\right)^p \left( k N^2 \Gamma_{av} \right)^{\ell} \left(\ell+p\right)!}
{\left(N \Gamma_{av} + 1 \right)^{2\ell+p+1} \left(\ell!\right)^2 p!}
\sum_{S_{q,\ell+p+1}}
\prod_{r=1}^{\ell+p+1} \frac{1}{q_r!} \nn \\
&& \times
\left[ \frac{2^{r-1}}{r}
\left[\frac{r\pi N^2 \Gamma_{av}}{4} L_{1/2}^2 \left(-k\right)
\left( N \Gamma_{av} \left(1+k-\frac{\pi}{4}L_{1/2}^2\left(-k\right)\right)
+ \frac{1}{2}\right)^{r-1} \right. \right. \nn \\
&& \hspace{2cm} \left. \left. + \left( N \Gamma_{av}
\left( 1+k-\frac{\pi}{4} L_{1/2}^2 \left(-k\right) \right) + \frac{1}{2} \right)^{r} 
+ \frac{1}{2^r} \right] \right]^{q_r} \, ,
\label{eq25}
\eeqarr
where $\Gamma_{av} \dn E_s \sigma_h^2/N_0$ is defined as the average SNR per receive diversity branch and the summation over the set $\sum_{S_{q,\ell+p+1}}$ is carried out for all the possible tuples of $q_1,\ldots,q_{\ell+p+1}$ such that $\sum_{r=1}^{\ell+p+1} rq_r = \ell+p+1$.

{\em Proof}: The proof is presented in Appendix I. \hfill $\blacksquare$
%%%%%%%%%%%%%%%%%%%%%%%%%%%%%%%%%%%%%%%%%%%%%%%%%%%%%%%%%%%%%%%%%%%%%%%%%%%%%%%%%%%%%%%%%%%%
%%%%%%%%%%%%%%%%%%%%%%%%%%%%%%%%%%%%%%%%%%%%%%%%%%%%%%%%%%%%%%%%%%%%%%%%%%%%%%%%%%%%%%%%%%%%
\vspace{-0.5cm}
\subsection{PED Analysis}
We now wish to compute the expression of the PED in a similar manner. Let $Y_{ssk_1}, \ldots ,Y_{ssk_L}$ denote the expression corresponding to (\ref{eq12}) and (\ref{eq9}) for all the non-target antennas at the receiver end with $L= N_{R_{X}}-1$. It is observed that $Y_{ssk_1}, \ldots ,Y_{ssk_L}$ are statistically i.i.d.. Thus, the probability of erroneous detection, denoted by $P_{e,ssk}$, utilizing the greedy detector in (\ref{eq5}), is expressed by
\beq
P_{e,ssk} = 1 - \Pr \left\{Y_{ssk_1}, \ldots ,Y_{ssk_L} < X_{ssk} \right\}
= 1 - \int\limits_{-\infty}^{\infty}
\left( \prod \limits_{i=1}^{L} F_{Y_{ssk_{i}}} \left(x \right) \right)
f_{X_{ssk}} \left(x\right) \textnormal{d}x .
\label{eq26}
\eeq
Furthermore, using similar steps as in (\ref{eq17}) (see Appendix I), we have
\beqarr
&& \! \! \! \! \! \! \! \! \! \! \! \! \! \! \! \!
\prod\limits_{i=1}^{L} F_{Y_{ssk_i}} \left(x \right)
= \int\limits_{0}^{x} \ldots \int\limits_{0}^{x}
\prod_{i=1}^L \left[
\frac{\exp \left\{-\frac{y_i+E_{s}N^2|\mu|^2}{N\sigma_{h}^2E_s + N_0}\right\}}
{N\sigma_{h}E_s + N_0}
I_0 \left(\frac{2\sqrt{E_{s}N^2|\mu|^2y_i}}{N\sigma_{h}^2 E_s + N_0}\right) \right]
\textnormal{d}y_1 \ldots \textnormal{d}y_L \nn \\
&& \! \! \! \! \! \! \! \! \! \! \! \! \! \! \! \!
\stackrel{(a)}{=}
e^{ - \frac{L E_s N^2 \left| \mu \right|^2}
{N E_s \sigma_h^2 + N_0}}
\sum_{\ell_1,\ldots, \ell_L=0}^{\infty}
\sum_{p_1,\ldots,p_L=0}^{\infty}
\frac{ \left(-1 \right)^{\sum\limits_{i=1}^L p_i}
\left( E_s N^2 \left| \mu \right|^2 \right)^{\sum\limits_{i=1}^L \ell_i}
x^{\sum\limits_{i=1}^L \left(\ell_i + p_i + 1 \right)}}
{ \left( N E_s \sigma_h^2 + N_0 \right)^{L + \sum\limits_{i=1}^L 2 \ell_i + p_i}
\prod\limits_{i=1}^L \left( \ell_i ! \right)^2 p_i ! \left( \ell_i + p_i + 1 \right)} , 
\label{eq27}
\eeqarr
where $(a)$ is obtained by using the series-form expansions of $I_0(x)$ and $\exp \left\{x \right\}$ in (\ref{eq18}).

Upon substituting (\ref{eq27}) in (\ref{eq26}) to obtain the expression of $P_{e,ssk}$, we finally need to solve the integration given as
\beqarr
\int_{0}^{\infty} \! \! \! x^{\sum\limits_{i=1}^L
\left( \ell_i + p_i + 1 \right)} f_{X_{ssk}} (x) \textnormal{d} x
\! \! \! \! &=& \! \! \! \! \undb{E} \left[ X_{ssk}^{\sum\limits_{i=1}^L
\left( \ell_i + p_i + 1 \right)} \right]
\! \! = \! \! \frac{1}{\jmath^{\sum\limits_{i=1}^L
\left( \ell_i + p_i + 1 \right)}} \left.
\frac{\partial^{\sum\limits_{i=1}^L
\left( \ell_i + p_i + 1 \right)}}
{\partial \omega^{\sum\limits_{i=1}^L
\left( \ell_i + p_i + 1 \right)}}
\Psi_{X_{ssk}} \left(\jmath \omega \right) 
\right|_{\omega=0} \! \! \! \! \! \! . 
\label{eq28}
\eeqarr
The expression in (\ref{eq28}) is similar to that obtained in (\ref{eq20}). Thus, following similar steps from (\ref{eq22})-(\ref{eq23}) and by using Faa-di Bruno's formula, the series-form expression for the probability of erroneous antenna detection $P_{e,ssk}$ is obtained as 
\beqarr
P_{e,ssk} \! \! \! \! &=& \! \! \! \!
1 - \exp \left\{ - \frac{ k L N^2 \Gamma_{av}}
{N \Gamma_{av} + 1} \right\}
\sum_{\ell_1,\ldots, \ell_L=0}^{\infty}
\sum_{p_1,\ldots,p_L=0}^{\infty}
\frac{ \left(-1 \right)^{\sum\limits_{i=1}^L p_i}
\left( k N^2 \Gamma_{av} \right)^{\sum\limits_{i=1}^L \ell_i}
\prod\limits_{i=1}^L \left(\ell_i + p_i \right)!}
{ \left( N \Gamma_{av} + 1 \right)^{\alpha_{\ell,p} + \sum\limits_{i=1}^L \ell_i}
\prod\limits_{i=1}^L \left( \ell_i ! \right)^2 p_i !} \nn \\
&\times& \! \! \! \! \sum_{S_{q,\alpha_{\ell,p}}}
\prod_{r=1}^{\alpha_{\ell,p}} \frac{1}{q_r!}
\left[ \frac{2^{r-1}}{r}
\left[\frac{r\pi N^2 \Gamma_{av}}{4} L_{1/2}^2 \left(-k\right)
\left( N \Gamma_{av} \left(1+k-\frac{\pi}{4}L_{1/2}^2\left(-k\right)\right)
+ \frac{1}{2}\right)^{r-1} \right. \right. \nn \\
&& \hspace{4.5cm} \left. \left. + \left( N \Gamma_{av}
\left( 1+k-\frac{\pi}{4} L_{1/2}^2 \left(-k\right) \right) + \frac{1}{2} \right)^{r} 
+ \frac{1}{2^r} \right] \right]^{q_r} ,
\label{eq29}
\eeqarr
where $\alpha_{\ell,p}=\sum_{i=1}^L \left( \ell_i + p_i + 1 \right)$ and the summation over the set $S_{q,\alpha_{\ell,p}}$ is carried out for all possible tuples of $q_1 \ldots, q_{\alpha_{\ell,p}}$ such that $\sum_{r=1}^{\ell+p+1} rq_r = \alpha_{\ell,p}$.
%%%%%%%%%%%%%%%%%%%%%%%%%%%%%%%%%%%%%%%%%%%%%%%%%%%%%%%%%%%%%%%%%%%%%%%%%%%%%%%%%%%%%%%%%%%%%%%
%%%%%%%%%%%%%%%%%%%%%%%%%%%%%%%%%%%%%%%%%%%%%%%%%%%%%%%%%%%%%%%%%%%%%%%%%%%%%%%%%%%%%%%%%%%%%%%
\vspace{-0.2cm}
\subsection{Asymptotic Analysis}
For the case of high average SNR per branch, i.e., $\Gamma_{av} \gg 1$, the expression of $P_{e,ssk}$ in (\ref{eq29}) simplifies to
\beqarr
&& \! \! \! \! \! \! \! \! \! \! \! \! \!
P_{e,ssk \big|_{\Gamma_{av} \gg 1}}
=
1 - e^{- k L N}
\sum_{\ell_1,\ldots, \ell_L=0}^{\infty}
\sum_{p_1,\ldots,p_L=0}^{\infty}
\left(-1 \right)^{\sum\limits_{i=1}^L p_i}
\left( k N\right)^{\sum\limits_{i=1}^L \ell_i}
\frac{\prod\limits_{i=1}^L \left(\ell_i + p_i \right)!}
{ \prod\limits_{i=1}^L \left( \ell_i ! \right)^2 p_i !} 
\sum_{S_{q,\alpha_{\ell,p}}}
\prod_{r=1}^{\alpha_{\ell,p}} \frac{1}{q_r!}
\nn \\
&&
\times
\left[ \frac{2^{r-1}}{r}
\left[\frac{r\pi N }{4} L_{1/2}^2 \left(-k\right)
\left(1+k-\frac{\pi}{4}L_{1/2}^2\left(-k\right) \right)^{r-1}
+ \left( 1+k-\frac{\pi}{4} L_{1/2}^2 \left(-k\right)\right)^{r} 
\right] \right]^{q_r} \! \! \! \! .
\label{eq30}
\eeqarr
It can be observed from (\ref{eq30}) that $P_{e,ssk \big|_{\Gamma_{av} \gg 1}}$ is independent of $\Gamma_{av}$ implying that the PED of the considered RIS-assisted SSK system under consideration saturates to the value obtained in (\ref{eq30}) at high $\Gamma_{av}$.
Similarly, for the case of the average SNR per branch being very small, i.e., $\Gamma_{av} \ll 1$, the PED in (\ref{eq29}) simplifies to
\vspace{-0.5cm}
\beq
P_{e,ssk \big|_{\Gamma_{av} \ll 1}} \! \! \! \! =
1 - e^{- k L N^2 \Gamma_{av}} \! \! \!
\sum_{\ell_1,\ldots, \ell_L=0}^{\infty}
%\ldots \sum_{\ell_L=0}^{\infty}
\sum_{p_1,\ldots,p_L=0}^{\infty}
%\ldots \sum_{p_L=0}^{\infty}
% \nn \\
% && \times
\frac{\left(-1 \right)^{\sum\limits_{i=1}^L p_i}
\left( k N^2 \Gamma_{av} \right)^{\sum\limits_{i=1}^L \ell_i}
\prod\limits_{i=1}^L \left(\ell_i + p_i \right)!}
{\prod\limits_{i=1}^L \left( \ell_i ! \right)^2 p_i !}
\sum_{S_{q,\alpha_{\ell,p}}}
% \hspace{-1.5cm} \sum_{\begin{array}{c}
% {\scriptstyle q_1 \ldots, q_{\alpha_{\ell,p}}} \vspace{-0.2cm} \\
% {\scriptstyle 0 \leq q_1,\ldots,q_{\alpha_{\ell,p}} \leq \alpha_{\ell,p} } \vspace{-0.2cm} \\
% {\scriptstyle q_1 + 2q_2 + \ldots + \alpha_{\ell,p} q_{\alpha_{\ell,p}} = \alpha_{\ell,p}}
% \end{array}} \hspace{-1.5cm}
\prod_{r=1}^{\alpha_{\ell,p}} \frac{1}{q_r! r^{q_r}} .
\label{eq31}
\eeq
%\end{small}
Finally, for the case of $\Gamma_{av}=0$, the PPED of the RIS-assisted wireless system antenna for the SSK scheme simplifies to
\vspace{-0.5cm}
%\beqarr
%P_{e,ssk \big|_{\Gamma_{av}=0}} \! \! \! \! &=& \! \! \! \!
%1 - \sum_{p_1=0}^{\infty} \ldots \sum_{p_L=0}^{\infty}
%\left(-1 \right)^{\sum\limits_{i=1}^L p_i} \nn \\
%&& \times \hspace{-2.5cm} \sum_{\begin{array}{c} {\scriptstyle q_1 \ldots, q_{\sum\limits_{i=1}^L
%\left( p_i + 1 \right)}} \\
%{\scriptstyle 0 \leq q_1,\ldots,q_{\sum\limits_{i=1}^L
%\left( p_i + 1 \right)} \leq \sum\limits_{i=1}^L
%\left( p_i + 1 \right) } \\
%{\scriptstyle q_1 + 2q_2 + \ldots + \left( \sum\limits_{i=1}^L
%\left( p_i + 1 \right) \right) q_{ \sum\limits_{i=1}^L
%\left( p_i + 1 \right) } = \sum\limits_{i=1}^L
%\left( p_i + 1 \right) }
%\end{array}} \hspace{-2cm}
%\prod_{r=1}^{\sum\limits_{i=1}^L
%\left( p_i + 1 \right)} \frac{1}{q_r! r^{q_r}} .
%\label{eq32}
%\eeqarr
\beq
P_{e,ssk \big|_{\Gamma_{av}=0}} =
1 - \sum_{p_1,\ldots,p_L=0}^{\infty}
%\ldots \sum_{p_L=0}^{\infty}
\left(-1 \right)^{\sum\limits_{i=1}^L p_i}
\sum_{S_{q,\sum_{i=1}^L \left(p_i +1 \right)}}
% \hspace{-2.5cm} \sum_{\begin{array}{c} {\scriptstyle q_1 \ldots, q_{\sum\limits_{i=1}^L
% \left( p_i + 1 \right)}} \\
% {\scriptstyle 0 \leq q_1,\ldots,q_{\sum\limits_{i=1}^L
% \left( p_i + 1 \right)} \leq \sum\limits_{i=1}^L
% \left( p_i + 1 \right) } \\
% {\scriptstyle q_1 + 2q_2 + \ldots + \left( \sum\limits_{i=1}^L
% \left( p_i + 1 \right) \right) q_{\sum\limits_{i=1}^L
% \left( p_i + 1 \right) } = \sum\limits_{i=1}^L
% \left( p_i + 1 \right) }
% \end{array}} \hspace{-2cm}
\prod_{r=1}^{\sum\limits_{i=1}^L
\left( p_i + 1 \right)} \frac{1}{q_r! r^{q_r}} ,
\label{eq32}
\eeq
where the summation over the set \hspace{-0.2cm} $\sum\limits_{S_{q,\sum_{i=1}^L \left(p_i +1 \right)}}$ \hspace{-0.3cm} is carried out for all possible tuples of $q_1 \ldots, q_{\sum\limits_{i=1}^L
\left( p_i + 1 \right)}$ such that $\sum_{r=1}^{\sum\limits_{i=1}^L
\left( p_i + 1 \right)} rq_r = \sum\limits_{i=1}^L
\left( p_i + 1 \right)$.
%%%%%%%%%%%%%%%%%%%%%%%%%%%%%%%%%%%%%%%%%%%%%%%%%%%%%%%%%%%%%%%%%%%%%%%%%%%%%%%%%%%%%%%%%%%%%%%
%%%%%%%%%%%%%%%%%%%%%%%%%%%%%%%%%%%%%%%%%%%%%%%%%%%%%%%%%%%%%%%%%%%%%%%%%%%%%%%%%%%%%%%%%%%%%%%
%%%%%%%%%%%%%%%%%%%%%%%%%%%%%%%%%%%%%%%%%%%%%%%%%%%%%%%%%%%%%%%%%%%%%%%%%%%%%%%%%%%%%%%%%%%%%%%
%%%%%%%%%%%%%%%%%%%%%%%%%%%%%%%%%%%%%%%%%%%%%%%%%%%%%%%%%%%%%%%%%%%%%%%%%%%%%%%%%%%%%%%%%%%%%%%
\section{PED Analysis of the RIS-Assisted SM System}
The performance evaluation of the RIS-SM system follows a similar approach to that of the RIS-SSK scheme.
%%%%%%%%%%%%%%%%%%%%%%%%%%%%%%%%%%%%%%%%%%%%%%%%%%%%%%%%%%%%%%%%%%%%%%%%%%%%%%%%%%%%%%%%%%%%%%%
%%%%%%%%%%%%%%%%%%%%%%%%%%%%%%%%%%%%%%%%%%%%%%%%%%%%%%%%%%%%%%%%%%%%%%%%%%%%%%%%%%%%%%%%%%%%%%%
\vspace{-0.5cm}
\subsection{PPED Analysis}
First, we denote two indices $w$ and $\hat{w}$ corresponding to the target and some other non-target receive diversity branch and compute the PPED given by
%\begin{small}
\beqarr
\Pr \left\{ \left| z_{w_{sm}} \right|^2 < \left| z_{\hat{w}_{sm}} \right|^2 \right\} \! \! \! \! &=& \! \! \! \!
\Pr \left\{ \left| \left( \sum_{i=1}^{N} h_{w,i} e^{\jmath \phi_{i}} \right) v + n_w \right|^2 
< \left|\left(\sum_{i=1}^{N} h_{\hat{w},i} e^{\jmath \phi_{i}} \right) v + n_{\hat{w}} \right|^2 \right\} \! ,
\label{33}
\eeqarr
%\end{small}
where $v$ is the data symbol belonging to the $M$-ary QAM/$M$-PSK constellation, $\undb{E} \left[|v|^2\right] = E_s$, which can also be expressed as
%\vspace{-0.2cm}
\beq
%\frac{v}{\sqrt{E_s}} 
v/\sqrt{E_s} = \Re \left\{v \right\} + \jmath \Im \left\{v \right\} \, .
\label{34}
%\vspace{-0.2cm}
\eeq
Similar to the case of the RIS-assisted SSK system, (\ref{33}) can be expressed in terms of an optimization problem to maximize the signal energy at the target antenna as
\beqarr
\min_{\left\{\phi_i\right\}_{i=1}^N} \! \! \! \! && \! \! \! \!
\text{Pr} \left\{ \left| \sum_{i=1}^{N} \beta_{\hat{w},i}
\exp \left\{\jmath \left( \phi_i - \psi_{\hat{w_i},i} \right) \right\}v\right|^2
> \left| \sum_{i=1}^{N} \beta_{w,i}
\exp \left\{\jmath \left( \phi_i-\psi_{w_i,i} \right) \right\}v \right|^2 \right\} . 
\label{eq35}
\eeqarr
Furthermore, the result in (\ref{eq35}) can be minimized by letting $\phi_i = \psi_{w,i}$ for $i=1, \ldots ,N$, resulting in the optimization problem leading to
% simplifying to
% \beq
% \hspace{-0.08cm}
% \min_{\left\{\phi_i\right\}_{i=1}^N} \! \!
% \text{Pr} \left\{ \left| \sum_{i=1}^{N} \beta_{\hat{w},i}
% \exp \left\{ \jmath \left( \phi_i-\psi_{\hat{w_i},i} \right) \right\}v \right|^2
% \!\!\!>\! \left| \sum_{i=1}^{N} \beta_{w,i}v \right|^2 \right\} ,
% \label{eq36}
% \eeq
% which further leads to
the expression of the PPED in (\ref{33}) being written as
%\begin{small}
\beqarr
\text{Pr} \left\{ \left| z_{w_{sm}} \right|^2 \! \! < \! \left| z_{\hat{w}_{sm}} \right|^2 \right\} 
= \text{Pr} \left\{ \left| \sum_{i=1}^{N} \beta_{\hat{w},i}
\exp \left\{ \jmath \left( \psi_{w,i} - \psi_{\hat{w},i} \right)\right\}v
+ n_{\hat{w}} \right|^2 
\! \! > \! \left| \sum_{i=1}^{N} \beta_{w,i}v + n_w \right|^2 \right\} \! .
\label{eq37}
\eeqarr
%\end{small}
% 2 column format
% \beqarr
% && \! \! \! \! \! \! \! \! \! \! \! \! \!\! \! \!\! \! \! 
% \text{Pr} \left\{ \left| z_{w_{sm}} \right|^2 < \left| z_{\hat{w}_{sm}} \right|^2 \right\} \nn \\
% && \! \! \! \!\!\!
% = \text{Pr} \left\{ \left| \sum_{i=1}^{N} \beta_{\hat{w},i}
% \exp \left\{ \jmath \left( \psi_{w,i} - \psi_{\hat{w},i} \right)\right\}v
% + n_{\hat{w}} \right|^2 \right. \nn \\
% && \hspace{3.2cm}  \left.
% > \left| \sum_{i=1}^{N} \beta_{w,i}v + n_w \right|^2 \right\} .
% \label{37}
% \eeqarr

To obtain the solution to (\ref{eq37}), we first find the statistics of the terms involved in (\ref{eq37}), which can be obtained by using the central limit theorem as
%\begin{small}
% 2 column format
% \beqarr
% && \hspace{-1.6cm} \sum_{i=1}^N \beta_{\hat{m},i}
% \exp \left\{ \jmath \left( \psi_{m,i}-\psi_{\hat{m},i} \right) \right\}v + n_{\hat{m}} \nn \\
% && \hspace{1.7cm} \sim {\mathcal{CN}} \left( N \sqrt{E_s} \mu \, , \, N E_{s} \sigma_{h}^2 + N_0 \right).
% \label{eq38}
% \eeqarr
\beq
\sum_{i=1}^N \beta_{\hat{m},i}
\exp \left\{ \jmath \left( \psi_{m,i}-\psi_{\hat{m},i} \right) \right\}v + n_{\hat{m}} \sim {\mathcal{CN}} \left( N \sqrt{E_s} \mu \, , \, N E_{s} \sigma_{h}^2 + N_0 \right),
\label{eq38}
\eeq

% 2 column format
% \beqarr
% \hspace{-0.1cm}\Re \left\{ \sum_{i=1}^{N} \beta_{m,i}v + n_m \right\}
% \!\sim\! {\mathcal{N}} 
% \left( \frac{N\sigma_{h} \sqrt{\pi E_s}\left(\Re \left(v \right)\right)}{2} L_\frac{1}{2} \left(-k \right), \right. \nn \\
% && \hspace{-7.5cm} \left. N E_s \sigma_{h}^2 \left(\Re \left(v \right)\right)^2
% \left( 1 + k - \frac{\pi}{4} L^2_\frac{1}{2} \left(-k \right) \right) + \frac{N_0}{2} \right) .
% \label{eq39}
% \eeqarr
%\begin{small}
\beqarr
\Re \left\{ \sum_{i=1}^{N} \beta_{m,i}v + n_m \right\}
\! \! \! \! & \sim & \! \! \! \! {\mathcal{N}} 
\left( \frac{N\sigma_{h} \sqrt{\pi E_s}\left(\Re \left\{v \right\}\right)}{2} L_\frac{1}{2} \left(-k \right), \right. \nn \\
&& \left. N E_s \sigma_{h}^2 \left(\Re \left\{v \right\}\right)^2
\left( 1 + k - \frac{\pi}{4} L^2_\frac{1}{2} \left(-k \right) \right) + \frac{N_0}{2} \right) ,
\label{eq39}
\eeqarr
%\end{small}
and
\beqarr
\hspace{-0.1cm}\Im \left\{ \sum_{i=1}^{N} \beta_{m,i}v + n_m \right\}
\!\sim\! {\mathcal{N}} 
\left( \frac{N\sigma_{h} \sqrt{\pi E_s}\left(\Im \left(v \right)\right)}{2} L_\frac{1}{2} \left(-k \right), \right. \nn \\
&& \hspace{-7.5cm} \left. N E_s \sigma_{h}^2 \left(\Im \left(v \right)\right)^2
\left( 1 + k - \frac{\pi}{4} L^2_\frac{1}{2} \left(-k \right) \right) + \frac{N_0}{2} \right) .
\label{eq40}
\eeqarr
% \beq
% \Im \left\{ \sum_{i=1}^{N} \beta_{m,i}v + n_m \right\}
% \!\sim\! {\mathcal{N}} 
% \left( \frac{N\sigma_{h} \sqrt{\pi E_s}\left(\Im \left\{v \right\} \right)}{2} L_\frac{1}{2} \left(-k \right),  N E_s \sigma_{h}^2 \left(\Im \left\{v \right\}\right)^2
% \left( 1 + k - \frac{\pi}{4} L^2_\frac{1}{2} \left(-k \right) \right) + \frac{N_0}{2} \right) .
% \label{eq40}
% \eeq
%\end{small}

Further, using the same approach as the analysis for the RIS-SSK system ($\left(8\text{a}\right)$ and $\left(8\text{b}\right)$), we define two random variables $X_{sm}$ corresponding to the instantaneous signal energy at the target antenna (related to $|z_{w_{sm}}|^2$) and $Y_{sm}$ corresponding to the instantaneous signal energy at any other non-target antenna (related to $|z_{\hat{w}_{sm}}|^2$), where $X_{sm}$ can be statistically modeled as
\beq
X_{sm} = \left|(V_a + V_1)+\jmath(V_b + V_2)\right|^2,
\label{eq41}
\eeq
with $V_a,V_b,V_1$, and $V_2$ being independent random variables, which can be statistically modeled using (\ref{eq38})-(\ref{eq40}) as
\bsub
%\begin{small}
\beq
V_1, V_2 \sim {\mathcal{N}} \left(0,c\right) \ , \
V_a \sim {\mathcal{N}} \left(\mu_1, b_1 \right) \ , \
V_b \sim {\mathcal{N}} \left(\mu_2, b_2 \right) \, ,
\label{eq42a}
\eeq
where
%\beqarr
%\mu_1 \!\!=\!\! \mu_x\Re \left(v\right),
%\mu_2 \!\!=\!\! \mu_x\Im \left(v\right),\nn \\
%\hspace{0.20cm} b_1 \!\!=\!\! b\Re \left(v \right)^2,
% b_2 \!\!=\!\! b \Im \left(v \right)^2 ,\
% \label{eq42b}
%\eeqarr
\beq
\mu_1 = \mu_X \Re \left\{v\right\} \ , \
\mu_2 = \mu_x\Im \left\{v \right\} \ , \
b_1 = b \left(\Re \left\{v \right\} \right)^2 \ , \
b_2 = b \left(\Im \left\{v \right\} \right)^2 \, ,
\label{eq42b}
\eeq
%\end{small}
\esub
and the variables $\mu_X$ and $b$ are given in (\ref{eq15}). Unlike the RIS-assisted SSK scheme, it is observed here the statistics of the terms involved in (\ref{eq37}) are dependent on the constellation of the transmitted symbol $v$. Moreover, similar to the case of the RIS-assisted SSK scheme, the c.f. of $X_{sm}$ can be written as
\beq
\psi_{X_{sm}}\left(\jmath \omega\right)= \frac{\text{exp}\left\{\frac{\jmath \omega \mu_1^2}{1-2\jmath\omega(b_1+c)}+\frac{\jmath \omega \mu_2^2}{1-2\jmath\omega(b_2+c)}\right\}}{(1-2\jmath\omega(b_1+c))^\frac{1}{2}(1-2\jmath\omega(b_2+c))^\frac{1}{2}} \, .
\label{43}
\eeq

The statistics thus obtained for $X_{sm}$ and $Y_{sm}$ are utilized to derive the expression of the PPED of the system given by the following theorem.

{\em Theorem 2}: The expression of the PPED for the RIS-assisted SM system under consideration is given by
%\begin{small}
\beqarr
&& \! \! \! \! \! \! \! \! \! \! \! \!
\text{Pr} \left\{X_{sm} < Y_{sm}\right\} = 
1 - e^{-\frac{k N^2 \Gamma_{av}}{N \Gamma_{av} + 1}}
\sum_{\ell=0}^{\infty} \sum_{p=0}^{\infty}
\frac{\left(-1\right)^p \left( k N^2 \Gamma_{av}  \right)^{\ell} \left(\ell+p\right)!}
{\left(N \Gamma_{av} + 1 \right)^{2\ell+p+1} \left(\ell!\right)^2 p!}
\sum_{S_{q,\ell+p+1}}
\prod_{r=1}^{\ell+p+1}  \frac{1}{q_r!}
\left[ \frac{2^{r-1}}{r} \right.
% \hspace{-1cm}
% \sum_{\begin{array}{c} {\scriptstyle q_1 \ldots, q_{\ell+p+1}} \vspace{-0.3cm} \\
% {\scriptstyle 0 \leq q_1,\ldots,q_{\ell+p+1} \leq \ell+p+1} \vspace{-0.3cm} \\
% {\scriptstyle q_1 + 2q_2 + \ldots + \left( \ell+p+1\right) q_{\ell+p+1} = \ell+p+1}
% \end{array}}
\nn \\
&& \! \! \! \! \! \! \! \! \! \! \! \! \times 
\left[\frac{r\pi N^2 \Gamma_{av}\left(\Re \left\{v \right\}\right)^2}{4} L_{1/2}^2 \left(-k\right)
\left( N \Gamma_{av} \left(\Re \left\{v \right\} \right)^2\left(1+k-\frac{\pi}{4}L_{1/2}^2\left(-k\right)\right)
\! + \frac{1}{2}\right)^{r-1} 
\! \! \! + \frac{r\pi N^2 \Gamma_{av}}{4} \right. \nn \\
&& \! \! \! \! \! \! \! \! \! \! \! \! \times \left(\Im \left\{ v \right\} \right)^2
L_{1/2}^2 \left(-k\right)
\left( N \Gamma_{av} \left(\Im \left\{ v \right\} \right)^2\left(1+k-\frac{\pi}{4}L_{1/2}^2\left(-k\right)\right)
+ \frac{1}{2}\right)^{r-1} 
\! + \left( N \Gamma_{av} \left(\Re \left\{v \right\} \right)^2 \right.
\nn \\ 
&& \! \! \! \! \! \! \! \! \! \! \! \!
\left. \left. \left. 
\times 
\left( 1+k-\frac{\pi}{4} L_{1/2}^2 \left(-k\right) \right) + \frac{1}{2} \right)^{r} 
+\left( N \Gamma_{av} \left(\Im \left\{ v\right\} \right)^2
\left( 1+k-\frac{\pi}{4} L_{1/2}^2 \left(-k\right) \right) + \frac{1}{2} \right)^{r} \right] \right]^{q_r} .
\label{eq51}
\eeqarr

{\em Proof}: The proof is presented in Appendix II. \hfill $\blacksquare$
%%%%%%%%%%%%%%%%%%%%%%%%%%%%%%%%%%%%%%%%%%%%%%%%%%%%%%%%%%%%%%%%%%%%%%%%%%%%%%%%%%%%%%%%%%%%%%%
%%%%%%%%%%%%%%%%%%%%%%%%%%%%%%%%%%%%%%%%%%%%%%%%%%%%%%%%%%%%%%%%%%%%%%%%%%%%%%%%%%%%%%%%%%%%%%%
\vspace{-0.5cm}
\subsection{PED Analysis}
We now wish to compute the expression of the PED in a similar manner. Let $Y_{sm_1}, \ldots ,Y_{sm_L}$ denote the expression corresponding to (\ref{eq37}) for all the $L$ non-target antennas at the receiver end with $L= N_{R_{X}}-1$. It can be observed that $Y_{sm_1}, \ldots ,Y_{sm_L}$ are statistically i.i.d.. Thus, the probability of erroneous detection, denoted by $P_{e,sm}$, utilizing the greedy detector in (\ref{eq5}), is expressed by
% \beqarr
% P_{e,sm} \! \! \! \! &=& \! \! \! \! 1 - \Pr \left\{Y_{sm_1}, \ldots ,Y_{sm_L} < X_{sm} \right\} \nn \\
% &=& \! \! \! \! 1 - \int\limits_{-\infty}^{\infty}
% \left( \prod \limits_{i=1}^{L} F_{Y_{sm_{i}}} \left(x \right) \right)
% f_{X_{sm}} \left(x\right) \textnormal{d}x .
% \label{eq52}
% \eeqarr
\beq
P_{e,sm} = 1 - \Pr \left\{Y_{sm_1}, \ldots ,Y_{sm_L} < X_{sm} \right\} 
= 1 - \int\limits_{-\infty}^{\infty}
\left( \prod \limits_{i=1}^{L} F_{Y_{sm_{i}}} \left(x \right) \right)
f_{X_{sm}} \left(x\right) \textnormal{d}x .
\label{eq52}
\eeq
Similar to the case of the RIS-assisted SSK system, the product term in (\ref{eq52}) can be computed as
\beqarr
&& \! \! \! \! \! \! \! \! \! \! \! \! \! \! \! \! \! \! \! \!
\prod\limits_{i=1}^{L} F_{Y_{sm_i}} \left(x \right)
= \int\limits_{0}^{x} \ldots \int\limits_{0}^{x} \prod_{i=1}^L
\left[\frac{\exp \left\{-\frac{y_i+E_{s}N^2|\mu|^2}{N\sigma_{h}^2E_s + N_0}\right\}}
{N\sigma_{h}E_s + N_0}
I_0 \left(\frac{2\sqrt{E_{s}N^2|\mu|^2y_i}}{N\sigma_{h}^2 E_s + N_0}\right) \right]
\textnormal{d}y_1 \ldots \textnormal{d}y_L \nn \\
% && \! \! \! \! \! \! \! \! \! \! \! \! \! \! \! \!
% \stackrel{(a)}{=}
% \exp \left\{ - \frac{L E_s N^2 \left| \mu \right|^2}
% {N E_s \sigma_h^2 + N_0} \right\} 
% \sum_{\ell_1=0}^{\infty} \ldots \sum_{\ell_L=0}^{\infty}
% \sum_{p_1=0}^{\infty} \ldots \sum_{p_L=0}^{\infty}
% \frac{ \left(-1 \right)^{\sum\limits_{i=1}^L p_i}
% \left( E_s N^2 \left| \mu \right|^2 \right)^{\sum\limits_{i=1}^L \ell_i}
% \prod\limits_{i=1}^L \int_{0}^{x} \left[ y_i^{\ell_i+p_i} \textnormal{d} y_i \right]}
% { \left( N E_s \sigma_h^2 + N_0 \right)^{L + \sum\limits_{i=1}^L 2 \ell_i + p_i}
% \prod\limits_{i=1}^L \left( \ell_i ! \right)^2 p_i !} \nn \\
&& \! \! \! \! \! \! \! \! \! \! \! \! \! \! \! \! \! \!
\stackrel{(a)}{=} e^{ - \frac{L E_s N^2 \left| \mu \right|^2}
{N E_s \sigma_h^2 + N_0}}
\sum_{\ell_1,\ell_L=0}^{\infty}
%\ldots \sum_{\ell_L=0}^{\infty}
\sum_{p_1,\ldots,p_L=0}^{\infty}
%\ldots \sum_{p_L=0}^{\infty}
\frac{ \left(-1 \right)^{\sum\limits_{i=1}^L p_i}
\left( E_s N^2 \left| \mu \right|^2 \right)^{\sum\limits_{i=1}^L \ell_i}
x^{\sum\limits_{i=1}^L \left(\ell_i + p_i + 1 \right)}}
{ \left( N E_s \sigma_h^2 + N_0 \right)^{L + \sum\limits_{i=1}^L 2 \ell_i + p_i}
\prod\limits_{i=1}^L \left( \ell_i ! \right)^2 p_i ! \left( \ell_i + p_i + 1 \right)} ,
\label{eq53}
\eeqarr
%\end{small}
%\hspace{-0.15cm}
where the step $(a)$ is obtained by using the series-form expansions of $I_0(x)$ and $\exp \left\{x \right\}$ in (\ref{eq18}). Substituting (\ref{eq53}) in (\ref{eq52}) followed by algebraic simplifications results in the series-form expression of $P_{e,sm}$ to be obtained as
\vspace{-0.5cm}
%\begin{small}
\beqarr
&& \! \! \! \! \! \! \! \! \! \!
P_{e_{sm}} = 1 - e^{ - \frac{ k L N^2 \Gamma_{av} }
{N \Gamma_{av} + 1}}
\sum_{\ell_1,\ell_L=0}^{\infty}
%\ldots \sum_{\ell_L=0}^{\infty}
\sum_{p_1,\ldots,p_L=0}^{\infty}
%\ldots \sum_{p_L=0}^{\infty}
\frac{ \left(-1 \right)^{\sum\limits_{i=1}^L p_i}
\left( k N^2 \Gamma_{av} \right)^{\sum\limits_{i=1}^L \ell_i}
\prod\limits_{i=1}^L \left(\ell_i + p_i \right)!}
{ \left( N \Gamma_{av} + 1 \right)^{L + \sum\limits_{i=1}^L 2 \ell_i + p_i}
\prod\limits_{i=1}^L \left( \ell_i ! \right)^2 p_i !}
\sum_{S_{q,\alpha_{\ell,p}}}
\prod_{r=1}^{\sum\limits_{i=1}^L
\alpha_{\ell,p}} \! \! \! \frac{1}{q_r!}
\left[ \frac{2^{r-1}}{r}  \right.
\nn \\
&& \! \! \! \! \! \! \! \! \times
% \hspace{-1.6cm} \sum_{\begin{array}{c} {\scriptstyle q_1 \ldots, q_{\alpha_{\ell,p}}} \vspace{-0.3cm} \\
% {\scriptstyle 0 \leq q_1,\ldots,q_{
% \alpha_{\ell,p}} \leq \alpha_{\ell,p}} \vspace{-0.3cm} \\
% {\scriptstyle q_1 + 2q_2 + \ldots + \left( \alpha_{\ell,p} \right) q_{ \alpha_{\ell,p}} \! \! = \alpha_{\ell,p}}
% \end{array}} \hspace{-1.8cm}
\left[\frac{r\pi N^2 \Gamma_{av}\left(\Re \left\{ v \right\} \right)^2}{4} L_{1/2}^2 \left(-k\right)
\left( N \Gamma_{av} \left(\Re \left\{ v \right\} \right)^2\left(1+k-\frac{\pi}{4}L_{1/2}^2\left(-k\right)\right)
\! + \frac{1}{2}\right)^{r-1} \right. \!
\! \! \! + \frac{r\pi N^2 \Gamma_{av}}{4}
\nn \\
&& \! \! \! \! \! \! \! \! \times
\left(\Im \left\{v \right\} \right)^2 L_{1/2}^2 \left(-k\right)
\left( N \Gamma_{av} \left(\Im \left\{ v \right\} \right)^2\left(1+k-\frac{\pi}{4}L_{1/2}^2\left(-k\right)\right)
+ \frac{1}{2}\right)^{r-1} 
+ \left( N \Gamma_{av} \left(\Re \left\{ v \right\} \right)^2 \right. \nn \\
&& \! \! \! \! \! \! \! \! \times
\left. \left. \left.
\left( 1+k-\frac{\pi}{4} L_{1/2}^2 \left(-k\right) \right) + \frac{1}{2} \right)^{r} 
+\left( N \Gamma_{av} \left(\Im \left\{ v\right\} \right)^2
\left( 1+k-\frac{\pi}{4} L_{1/2}^2 \left(-k\right) \right) + \frac{1}{2} \right)^{r} \right] \right]^{q_r} .
\label{eq54}
\eeqarr
%\end{small}
% For the case of the channels following Rayleigh distributions, i.e., for $k=0$. the expression of the PED in (\ref{eq54}) simplifies to \cite[Eq. 24]{MaBhPa:14}.
%%%%%%%%%%%%%%%%%%%%%%%%%%%%%%%%%%%%%%%%%%%%%%%%%%%%%%%%%%%%%%%%%%%%%%%%%%%%% 
%%%%%%%%%%%%%%%%%%%%%%%%%%%%%%%%%%%%%%%%%%%%%%%%%%%%%%%%%%%%%%%%%%%%%%%%%%%%%
%\vspace{-0.6cm}
\subsection{Asymptotic Analysis}
For the case of low average SNR, i.e. $\Gamma_{av} \ll 1$, the expression of $P_{e,sm}$ in (\ref{eq54}) simplifies to
\vspace{-0.3cm}
%\begin{small}
\beqarr
P_{e_{sm} \big|_{\Gamma_{av}\ll 1}} \! \! \! \! \! \! &=& \! \! \! \!
1 - \left\{ e^{- k L N}  \right\}
\! \! \sum_{\ell_1,\ldots,\ell_L=0}^{\infty}
%\ldots \sum_{\ell_L=0}^{\infty}
\sum_{p_1,\ldots,p_L=0}^{\infty}
%\ldots \sum_{p_L=0}^{\infty}
\left(-1 \right)^{\sum\limits_{i=1}^L p_i} 
\frac{ \left( k N^2 \Gamma_{av}\right)^{\sum\limits_{i=1}^L \ell_i}\prod\limits_{i=1}^L \left(\ell_i + p_i \right)!}
{ \prod\limits_{i=1}^L \left( \ell_i ! \right)^2 p_i !} 
\sum_{S_{q,\alpha_{\ell,p}}}
\prod_{r=1}^{\alpha_{\ell,p}} \frac{1}{q_r!}
\nn \\
&\times&
% \hspace{-1.8cm}
% \sum_{\begin{array}{c} {\scriptstyle q_1 \ldots, q_{\alpha_{\ell,p}}} \vspace{-0.2cm} \\
% {\scriptstyle 0 \leq q_1,\ldots,q_{\alpha_{\ell,p}} \leq {\alpha_{\ell,p}}} \vspace{-0.2cm} \\
% {\scriptstyle q_1 + 2q_2 + \ldots + \left( {\alpha_{\ell,p}}\right) q_{\alpha_{\ell,p}} = {\alpha_{\ell,p}}}
% \end{array}} \hspace{-1.6cm}
\left[ \frac{1}{r}
\left[\left(\frac{r\pi N^2 \Gamma_{av} \left(\Re \left\{ v \right\} \right)^2}{4} L_\frac{1}{2}^2 \left(-k\right)\right) 
\! \! + \! \! 
\left(\frac{r\pi N^2 \left(\Im \left\{ v\right\} \right)^2}{4} L_{1/2}^2 \left(-k\right)\right) + 1 \right] \right]^{q_r} \! \! \! \! .
\label{eq56}
\eeqarr
%\end{small}
% Double-column format
%\begin{small}
%\beqarr
%&& \hspace{-8.4cm}
%P_{e_{sm},\Gamma_{av}\ll 1}  \nn \\
%\hspace{-5.2cm} \! \!   = \! \! 1 -  \left\{ e^{- k L N}  \right\}
%\sum_{\ell_1=0}^{\infty} \ldots \sum_{\ell_L=0}^{\infty}
%\sum_{p_1=0}^{\infty} \ldots \sum_{p_L=0}^{\infty} \left(-1 \right)^{\sum\limits_{i=1}^L p_i} \nn \\ &&   \hspace{-6cm} \times \frac{ 
%\left( k N^2 \Gamma_{av}\right)^{\sum\limits_{i=1}^L \ell_i}\prod\limits_{i=1}^L \left(\ell_i + p_i \right)!}
%{ \prod\limits_{i=1}^L \left( \ell_i ! \right)^2 p_i !} \nn \\
%&& \hspace{-8.7cm} 
%\sum_{\begin{array}{c} {\scriptstyle q_1 \ldots, q_{\alpha_{\ell,p}}} \\
%{\scriptstyle 0 \leq q_1,\ldots,q_{\alpha_{\ell,p}} \leq {\alpha_{\ell,p}}} \\
%{\scriptstyle q_1 + 2q_2 + \ldots + \left( {\alpha_{\ell,p}}\right) q_{\alpha_{\ell,p}} = {\alpha_{\ell,p}}}
%\end{array}} \hspace{-1.8cm}
%\prod_{r=1}^{\alpha_{\ell,p}} \frac{1}{q_r!}
%\left[ \frac{1}{r}
%\left[\left(\frac{r\pi N^2 \Gamma_{av} \left(\Re(v)\right)^2}{4} L_\frac{1}{2}^2 \left(-k\right)\right) \right. \right. \nn \\ \left. \left. \hspace{1.6cm} +\left(\frac{r\pi N^2 \left(\Im(v)\right)^2}{4} L_{1/2}^2 \left(-k\right)\right) + 1
% \right] \right]^{q_r} .\nn \\
%\label{eq56}
%\eeqarr
%\end{small}
Similarly for the case of high average SNR, i.e. $\Gamma_{av} \gg 1$, $P_{e,sm}$ in (\ref{eq54}) simplifies as
%\begin{small}
\beqarr
&& \! \! \! \! \!
P_{e_{sm} \big|_{\Gamma_{av}\gg 1}} \! \! \! =
1 - \exp \left\{ - k L N \right\}
\sum_{\ell_1,\ldots,\ell_L=0}^{\infty}
%\ldots \sum_{\ell_L=0}^{\infty}
\sum_{p_1,\ldots,p_L=0}^{\infty}
%\ldots \sum_{p_L=0}^{\infty}
\frac{ \left(-1 \right)^{\sum\limits_{i=1}^L p_i}
\left( k N^2 \right)^{\sum\limits_{i=1}^L \ell_i}
\prod\limits_{i=1}^L \left(\ell_i + p_i \right)!}
{ \left( N\right)^{L + \sum\limits_{i=1}^L 2 \ell_i + p_i}
\prod\limits_{i=1}^L \left( \ell_i ! \right)^2 p_i !} 
\nn \\
&&
% \hspace{-1.8cm} \sum_{\begin{array}{c} {\scriptstyle q_1 \ldots, q_{\alpha_{\ell,p}}} \vspace{-0.2cm} \\
% {\scriptstyle 0 \leq q_1,\ldots,q_{
% \alpha_{\ell,p}} \leq \alpha_{\ell,p}} \vspace{-0.2cm} \\
% {\scriptstyle q_1 + 2q_2 + \ldots + \left( \alpha_{\ell,p} \right) q_{ \alpha_{\ell,p}} = \alpha_{\ell,p}}
% \end{array}} \hspace{-1.8cm}
\times \sum_{S_{q,\alpha_{\ell,p}}}
\prod_{r=1}^{\alpha_{\ell,p}} \frac{1}{q_r!}
\left[ \frac{2^{r-1}}{r}
\left[\frac{r\pi N^2 \left(\Re \left\{ v \right\} \right)^2}{4} L_{1/2}^2 \left(-k\right)
\left( N  \left(\Re \left\{ v \right\} \right)^2\left(1+k-\frac{\pi}{4}L_{1/2}^2\left(-k\right)\right)
\right)^{r-1} \right. \right. \nn \\
&& \! \! \! \! \! \left. \left. +\frac{r\pi N^2 \left(\Im \left\{ v \right\} \right)^2}{4} L_{1/2}^2 \left(-k \right)
\left( N  \left(\Im \left\{ v\right\} \right)^2\left(1+k-\frac{\pi}{4}L_{1/2}^2\left(-k\right)\right)
\right)^{r-1} \right. \right. \nn \\
&& \! \! \! \! \! \left. \left. + \left( N  \left(\Re \left\{ v \right\} \right)^2
\left( 1+k-\frac{\pi}{4} L_{1/2}^2 \left(-k\right) \right) \right)^{r} 
+\left( N  \left(\Im \left\{ v\right\} \right)^2
\left( 1+k-\frac{\pi}{4} L_{1/2}^2 \left(-k\right) \right)  \right)^{r} \right] \right]^{q_r} .
\label{eq55}
\eeqarr

Similar to the case of the RIS-assisted SSK system, it can be observed from (\ref{eq56}) that the expression of $P_{e,sm \big|_{\Gamma_{av} \gg 1}}$ is independent of $\Gamma_{av}$ implying that the PED of the considered RIS-assisted SM system under consideration saturates to the value obtained in (\ref{eq56}) at high $\Gamma_{av}$.

Additionally for the case of $\Gamma_{av}=0$, the expression of the PED in (\ref{eq54}) is obtained as
\beq
P_{e,sm_{\Gamma_{av}=0}} =
1 - \sum_{p_1,\ldots,p_L=0}^{\infty}
%\ldots \sum_{p_L=0}^{\infty}
\left(-1 \right)^{\sum\limits_{i=1}^L p_i}
\sum_{S_{q,\sum_{i=1}^L \left(p_i +1 \right)}}
% \hspace{-2.5cm} \sum_{\begin{array}{c} {\scriptstyle q_1 \ldots, q_{\sum_{i=1}^L
% \left( p_i + 1 \right)}} \\
% {\scriptstyle 0 \leq q_1,\ldots,q_{\sum_{i=1}^L
% \left( p_i + 1 \right)} \leq \sum\limits_{i=1}^L
% \left( p_i + 1 \right) } \\
% {\scriptstyle q_1 + 2q_2 + \ldots + \left( \sum\limits_{i=1}^L
% \left( p_i + 1 \right) \right) q_{ \sum\limits_{i=1}^L
% \left( p_i + 1 \right) } = \sum_{i=1}^L
% \left( p_i + 1 \right) }
% \end{array}} \hspace{-2cm}
\prod_{r=1}^{\sum\limits_{i=1}^L
\left( p_i + 1 \right)} \frac{1}{q_r! r^{q_r}} \, .
\label{eq57}
\eeq
%Double-column format
%\beqarr
%P_{e,sm_{\Gamma_{av}=0}} \! \! \! \! &=& \! \! \! \!
%1 - \sum_{p_1=0}^{\infty} \ldots \sum_{p_L=0}^{\infty}
%\left(-1 \right)^{\sum\limits_{i=1}^L p_i} \nn \\
%&& \times \hspace{-2.5cm} \sum_{\begin{array}{c} {\scriptstyle q_1 \ldots, q_{\sum\limits_{i=1}^L
%\left( p_i + 1 \right)}} \\
%{\scriptstyle 0 \leq q_1,\ldots,q_{\sum\limits_{i=1}^L
%\left( p_i + 1 \right)} \leq \sum\limits_{i=1}^L
%\left( p_i + 1 \right) } \\
%{\scriptstyle q_1 + 2q_2 + \ldots + \left( \sum\limits_{i=1}^L
%\left( p_i + 1 \right) \right) q_{ \sum\limits_{i=1}^L
%\left( p_i + 1 \right) } = \sum\limits_{i=1}^L
%\left( p_i + 1 \right) }
%\end{array}} \hspace{-2cm}
%\prod_{r=1}^{\sum\limits_{i=1}^L
%\left( p_i + 1 \right)} \frac{1}{q_r! r^{q_r}} ,
%\label{eq57}
%\eeqarr
It can be noted by comparing (\ref{eq57}) and (\ref{eq32}) that the performance of the RIS-assisted SSK and the SM systems are exactly the same at $\Gamma_{av}=0$. Intuitively, this is an expected result as the SSK scheme differs from the SM scheme in terms of the constellation utilized for data transmission, and for $\Gamma_{av}=0$, the transmitted energy is zero.
%%%%%%%%%%%%%%%%%%%%%%%%%%%%%%%%%%%%%%%%%%%%%%%%%%%%%%%%%%%%%%%%%%%%%%%%%%%%%
%%%%%%%%%%%%%%%%%%%%%%%%%%%%%%%%%%%%%%%%%%%%%%%%%%%%%%%%%%%%%%%%%%%%%%%%%%%%%
%%%%%%%%%%%%%%%%%%%%%%%%%%%%%%%%%%%%%%%%%%%%%%%%%%%%%%%%%%%%%%%%%%%%%%%%%%%%%
%%%%%%%%%%%%%%%%%%%%%%%%%%%%%%%%%%%%%%%%%%%%%%%%%%%%%%%%%%%%%%%%%%%%%%%%%%%%%
\vspace{-0.5cm}
\section{SEP Analysis of the RIS-Assisted System}
In this section, we derive the expressions of the SEP of the RIS-assisted system for the cases of the transmitter utilizing the $M$-ary QAM and $M$-PSK constellations for data transmission. The meta-surfaces of the RIS are considered to introduce phase shifts on the transmitted signal such that the instantaneous SNR at the receiver is maximized, as given in (\ref{eq3}). In the case of a large number of meta-surfaces, i.e., $N \gg 1$, the square root of the maximum instantaneous SNR, denoted by $\gamma$ follows a Gaussian distribution as
%\begin{small}
\beq
\sqrt{\gamma} \sim {\mathcal{N}} \left(\frac{N L_{1/2}(-k)\sqrt{\Gamma_{av}\pi}}{2}, N \Gamma_{av}(1+k-\frac{\pi}{4}L^2_{1/2}(-k)) \right) \, .
\label{eq58}
%\label{eq41a}
\eeq
%\end{small}
It is to be noted that the distribution of $\gamma$ will be the same as $ \gamma_{w,max} \, , \forall w \in \left\{1,\ldots, N_{R_{\text{X}}} \right\}$. Furthermore, from (\ref{eq58}), it is observed that $\gamma$ follows a non-central $\chi^2(1)$ distribution and thus, its characteristic function is given by
\beq
\Psi_{\gamma}\left(\jmath \omega\right) = \frac{\text{exp}\left\{\frac{\frac{\jmath \omega N^2 \pi \Gamma_{av} L_{1/2}^2(-k)}{4}}{1-2\jmath \omega N \Gamma_{av}(1+k-\frac{\pi}{4}L_{1/2}^2(-k))}\right\}}{\left(1-2\jmath \omega N \Gamma_{av} \left(1+k-\frac{\pi}{4}L_{1/2}^2(-k) \right)\right)^\frac{1}{2}} \, .
\label{eq59}
\eeq
%%%%%%%%%%%%%%%%%%%%%%%%%%%%%%%%%%%%%%%%%%%%%%%%%%%%%%%%%%%%%%%%%%%%%%%%%%%%%
%%%%%%%%%%%%%%%%%%%%%%%%%%%%%%%%%%%%%%%%%%%%%%%%%%%%%%%%%%%%%%%%%%%%%%%%%%%%%
\subsection{SEP for $M$-PSK Constellation}
Using the c.f. of $\gamma$ in (\ref{eq59}), the expression for the SEP of the RIS-assisted system employing $M$-PSK constellation for data modulation can be expressed as \cite{vetu2910} 
\beq
P_{s,M\text{-PSK}}=\frac {1}{\pi} \int _{0}^{\frac{M-1}{M}\pi} \Psi_{\gamma } \left ({\frac {-\sin^2 \frac{\pi}{M}}{\sin ^{2} \! \eta }}\right) \text{d} \eta \, .
\label{eq60}
\eeq
Substituting (\ref{eq59}) in (\ref{eq60}), the expression for the SEP can be obtained by solving the expression as
\beq
P_{s,M\text{-PSK}} = \frac {1}{\pi} \int _{0}^{\frac{M-1}{M}\pi}
\frac{\text{exp}\left\{-\frac{\frac{\frac {\sin^2 \frac{\pi}{M}}{\sin ^{2} \! \eta } N^2 \pi \Gamma_{av} L_{1/2}^2(-k)}{4}}{1+2\frac {\sin^2 \frac{\pi}{M}}{\sin ^{2} \! \eta } N \Gamma_{av}(1+k-\frac{\pi}{4}L_{1/2}(-k))}\right\} d\eta}{\left(1+2\frac {\sin^2 \frac{\pi}{M}}{\sin ^{2} \! \eta } N \Gamma_{av}(1+k-\frac{\pi}{4}L_{1/2}^
2(-k))\right)^\frac{1}{2}} \, .
\label{eq61}
\eeq

For the case of high average SNR, i.e., $\Gamma_{av} \gg 1$, we have
\beqarr
1+2\frac {\sin^2 \frac{\pi}{M}}{\sin ^{2} \! \eta } N \Gamma_{av}
\left( 1+k-\frac{\pi}{4}L_{1/2}^2(-k) \right)
\approx
2\frac {\sin^2 \frac{\pi}{M}}{\sin ^{2} \! \eta } N \Gamma_{av}
\left(1+k-\frac{\pi}{4}L_{1/2}^2(-k) \right) \, ,
\eeqarr
using which the asymptotic expression of the SEP at high $\Gamma_{av}$ utilizing algebraic simplification is obtained as 
\beqarr
P_{s,M\text{-PSK}\big|_{\Gamma_{av} \gg 1}} \! \! \! \! &=& \! \! \! \!
\frac{1-\cos \left(\frac{M-1}{M} \right)\pi}
{\pi \left(2N\Gamma_{av} \left(1+k-\frac{\pi}{4} L_{1/2}^2 \left(-k\right) \right)\right)^\frac{1}{2}} 
\exp \left\{\frac{-N \pi L_{1/2}^2 \left(-k\right)}{8 \left( 1+k-\frac{\pi}{4} L_{1/2}^2 \left(-k\right) \right)}\right\}.
\label{eq62}
\eeqarr

For the case of $\Gamma_{av} \ll 1$ and $M=2$ we have
\beq
1+2\frac {\sin^2 \frac{\pi}{M}}{\sin ^{2} \! \eta } N \Gamma_{av}
\left(1+k-\frac{\pi}{4}L_{1/2}^2(-k) \right) \approx 1
\label{eq63}
\eeq
which, when substituted in (\ref{eq61}) followed by algebraic simplifications, results in the expression of the SEP to be obtained as
\beq
P_{s,M\text{-PSK} \big|_{\Gamma_{av} \ll 1,M=2}} \approx Q \left(\sqrt{\frac{N^2 \pi \Gamma_{av}L^2_{1/2}(-k)}{2}} \right) \, ,
\label{eq64}
\eeq
where $Q \left( \cdot \right)$ denotes the Gaussian-$Q$ function.
Additionally, for the case of $M>2$ the corresponding approximation becomes
% \beqarr
% && \!\!\!\!\!\!\!\!\!\!\!
% \frac{\text{exp}\left\{-\frac{\frac{\frac {\sin^2 \frac{\pi}{M}}{\sin ^{2} \! \eta } N^2 \pi \Gamma_{av} L_{1/2}^2(-k)}{4}}{1+2\frac {\sin^2 \frac{\pi}{M}}{\sin ^{2} \! \eta } N \Gamma_{av}(1+k-\frac{\pi}{4}L_{1/2}(-k))}\right\}}{\left(1+2\frac {\sin^2 \frac{\pi}{M}}{\sin ^{2} \! \eta } N \Gamma_{av}(1+k-\frac{\pi}{4}L_{1/2}^
% 2(-k))\right)^\frac{1}{2}} \nn \\
% && \!\!\!\!\!\!\!\!\!\! \approx
% \text{exp}\!\left\{-\frac{\frac{\frac {\sin^2 \frac{\pi}{M}}{\sin ^{2} \! \eta } N^2 \pi \Gamma_{av} L_{1/2}^2(-k)}{4}}{1+2\frac {\sin^2 \frac{\pi}{M}}{\sin ^{2} \! \eta } N \Gamma_{av}(1+k-\frac{\pi}{4}L_{1/2}(-k))}\!\!\right\} \, ,
% \label{eq65}
% \eeqarr
\beq
\frac{\text{exp}\left\{-\frac{\frac{\frac {\sin^2 \frac{\pi}{M}}{\sin ^{2} \! \eta } N^2 \pi \Gamma_{av} L_{1/2}^2(-k)}{4}}{1+2\frac {\sin^2 \frac{\pi}{M}}{\sin ^{2} \! \eta } N \Gamma_{av}(1+k-\frac{\pi}{4}L_{1/2}(-k))}\right\}}{\left(1+2\frac {\sin^2 \frac{\pi}{M}}{\sin ^{2} \! \eta } N \Gamma_{av}(1+k-\frac{\pi}{4}L_{1/2}^
2(-k))\right)^\frac{1}{2}}
\approx
\text{exp}\!\left\{-\frac{\frac{\frac {\sin^2 \frac{\pi}{M}}{\sin ^{2} \! \eta } N^2 \pi \Gamma_{av} L_{1/2}^2(-k)}{4}}{1+2\frac {\sin^2 \frac{\pi}{M}}{\sin ^{2} \! \eta } N \Gamma_{av}(1+k-\frac{\pi}{4}L_{1/2}(-k))}\!\!\right\} \, ,
\label{eq65}
\eeq
which, when substituted in (\ref{eq61}) followed by algebraic simplifications, results in the asymptotic expression of the SEP to be given by
\beqarr
&& \!\!\!\!\!\!\!\!\!
P_{s, M\text{-PSK} \big|_{\Gamma_{av} \ll 1,M>2}} \approx
\frac{\left(M-1\right)}{M}
- \frac{N L_{1/2}^2(-k)}
{8 \left(1+k-\frac{\pi}{4}L_{1/2}^2(-k)\right) \sqrt{1+\zeta_M}} \nn \\
&&
\times
\left[\tan^{-1}\left(\sqrt{1 + \zeta_M}
\tan \left( \frac{\left(M-1\right)\pi}{M} \right) \right)
+\frac{\sqrt{1 + \zeta_M}}{2} \tan \left( \frac{\left(M-1\right)\pi}{M} \right) \right] \, .
\label{eq66}
\eeqarr
where
\beq
\zeta_M = \frac{1}{2 \sin^2 \frac{\pi}{M}N \Gamma_{av}
\left(1 + k -\frac{\pi}{4}L_{1/2}^2(-k)\right)} \, .
\label{eq67}
\eeq
%%%%%%%%%%%%%%%%%%%%%%%%%%%%%%%%%%%%%%%%%%%%%%%%%%%%%%%%%%%%%%%%%%%%%%%%%%%%%
%%%%%%%%%%%%%%%%%%%%%%%%%%%%%%%%%%%%%%%%%%%%%%%%%%%%%%%%%%%%%%%%%%%%%%%%%%%%%
\subsection{SEP for $M$-QAM Constellation}
Using the c.f. of $\gamma$ in (\ref{eq59}), the expression for the SEP of the RIS-assisted system employing $M$-ary QAM constellation for data modulation can be expressed as
\beqarr
P_{s , M\text{-QAM}} \! \! \! \! &=& \! \! \! \!
\frac{4}{\pi} \left (1-\frac{1}{\sqrt{M}}\right)
\int_{0}^{\frac{\pi}{2}} \! \! \Psi_{\gamma}
\left(\frac {-3}{2(M-1) \sin ^{2} \eta}\right) \text{d} \eta \nn \\
&& - \frac {4}{\pi} \left (1-\frac {1}{\sqrt{M}}\right)^2 
\int_{0}^{\frac{\pi}{4}} \! \! \Psi_{\gamma}
\left (\frac {-3}{2(M-1) \sin ^{2} \eta }\right) \text{d} \eta \, ,
\label{eq69}
\eeqarr
which using (\ref{eq59}) can be re-written as
%\begin{figure*}
%\begin{small}
\beqarr
P_{s , M\text{-QAM}} \! \! \! \! &=& \! \! \! \! 
\frac{4}{\pi} \left (1-\frac{1}{\sqrt{M}} \right)
\int_{0}^{\frac{\pi}{2}}
\frac{\text{exp}\left\{\frac{-\frac{3 N^2 \Gamma_{av}\pi }{8\left(M-1\right)\sin^2 \eta} L_{1/2}^2(-k)}{1+\frac{3 N \Gamma_{av}}{\left(M-1\right)\sin^2 \eta}
\left( 1+k-\frac{\pi}{4}L_{1/2}^2(-k) \right)}\right\}}
{\left(1+\frac{3 N \Gamma_{av}}{\left(M-1\right)\sin^2 \eta}
\left(1+k-\frac{\pi}{4}L_{1/2}^2(-k) \right)\right)^\frac{1}{2}} \nn \\
&& - \frac{4}{\pi} \left (1-\frac{1}{\sqrt{M}}\right)^2
\int_{0}^{\frac{\pi}{4}}
\frac{\text{exp}\left\{\frac{-\frac{3 N^2 \Gamma_{av}\pi }
{8\left(M-1\right)\sin^2 \eta} L_{1/2}^2(-k)}
{1+\frac{3 N \Gamma_{av}}{\left(M-1\right)\sin^2 \eta}
\left(1+k-\frac{\pi}{4}L_{1/2}^2(-k) \right)}\right\}}
{\left(1+\frac{3 N \Gamma_{av}}{\left(M-1\right)\sin^2 \eta}
\left(1+k-\frac{\pi}{4}L_{1/2}^2(-k) \right) \right)^\frac{1}{2}} \, .
\label{eq70}
\eeqarr
%\end{small}
%\noindent\rule{\textwidth}{.5pt}
%\end{figure*}

For the case of high average SNR values, i.e., $\Gamma_{av} \gg 1$, the terms in (\ref{eq70}) can be simplified using the approximation
% \beqarr
% &&
% \!\!\!\!\!\!\!\!\!\!\!\!\!\!\!\!\!\!\!\!\!
% 1+\frac{3 N \Gamma_{av}}{\left(M-1\right)\sin^2 \eta}
% \left(1+k-\frac{\pi}{4}L_{1/2}^2(-k) \right) \nn \\
% &&\qquad \approx
% \frac{3 N \Gamma_{av}}{\left(M-1\right)\sin^2 \eta}
% \left(1+k-\frac{\pi}{4}L_{1/2}^2(-k) \right) \, ,
% \label{eq71}
% \eeqarr
\beq
1+\frac{3 N \Gamma_{av}}{\left(M-1\right)\sin^2 \eta}
\left(1+k-\frac{\pi}{4}L_{1/2}^2(-k) \right)
\approx
\frac{3 N \Gamma_{av}}{\left(M-1\right)\sin^2 \eta}
\left(1+k-\frac{\pi}{4}L_{1/2}^2(-k) \right) \, ,
\label{eq71}
\eeq
using which, followed by algebraic simplifications, the asymptotic expression of the SEP is obtained as
% Double-column format
%\beqarr
%P_{s,M\text{-QAM}} \! \! \! \! &=& \! \! \! \!
%\frac {4}{\pi } \left ({1-\frac {1}{\sqrt {M}} }\right)\frac{\sqrt{2(M-1)}}{\sqrt{6N\Gamma_{av} \left(1+k-\frac{\pi}{4} L_{1/2}^2 \left(-k\right) \right)}} \nn \\ 
%&& \times \exp \left\{\frac{-N \pi L_{1/2}^2 \left(-k\right)}{8 \left(1+k-\frac{\pi}{4} L_{1/2}^2 \left(-k\right) \right)}\right\}\nn \\
%&-& \!\!\!\! \frac{4}{\pi} \left({1-\frac {1}{\sqrt {M}} }\right)^2\frac{\sqrt{\left(\sqrt{2}-1\right)(M-1)}}{\sqrt{6N\Gamma_{av} \left(1+k-\frac{\pi}{4} L_{1/2}^2 \left(-k\right) \right)}} \nn \\
%&& \times \exp \left\{\frac{-N \pi L_{1/2}^2. \left(-k\right)}{8 \left(1+k-\frac{\pi}{4} L_{1/2}^2 \left(-k\right) \right)}\right\} .
%\label{eq72}
%\eeqarr
%\begin{small}
\beqarr
P_{s,M\text{-QAM} \big|_{\Gamma_{av} \gg 1}} \! \! \! \! &=& \! \! \! \!
\frac {4}{\pi } \left ({1-\frac {1}{\sqrt {M}} }\right)\frac{\sqrt{2(M-1)}}{\sqrt{6N\Gamma_{av} \left(1+k-\frac{\pi}{4} L_{1/2}^2 \left(-k\right) \right)}} 
e^{\frac{-N \pi L_{1/2}^2 \left(-k\right)}{8 \left(1+k-\frac{\pi}{4} L_{1/2}^2 \left(-k\right) \right)}}\nn \\
&-& \!\!\!\! \frac{4}{\pi} \left({1-\frac {1}{\sqrt {M}} }\right)^2\frac{\sqrt{\left(\sqrt{2}-1\right)(M-1)}}{\sqrt{6N\Gamma_{av} \left(1+k-\frac{\pi}{4} L_{1/2}^2 \left(-k\right) \right)}}
e^{\frac{-N \pi L_{1/2}^2. \left(-k\right)}{8 \left(1+k-\frac{\pi}{4} L_{1/2}^2 \left(-k\right) \right)}} .
\label{eq72}
\eeqarr
%\end{small}
For the case of $\Gamma_{av} \ll 1$ the SEP expression can be simplified by approximating the first and second term of (\ref{eq70}) as
\beq
1+\frac{3 N \Gamma_{av}}{\left(M-1\right)\sin^2 \eta}
\left( 1+k-\frac{\pi}{4}L_{1/2}^2 \left(-k \right) \right) \approx 1 \, ,
\label{eq73}
\eeq
and
%\beqarr
%&& \!\!\!\!\!\!\!\!\!\!\!\!\!\!\!\!\!\!\!\!\!\!\!
%\frac{\text{exp}\left\{\frac{-\frac{3 N^2 \Gamma_{av}\pi }{8\left(M-1\right)\sin^2 \eta}  L_{1/2}^2(-k)}{1+\frac{3 N \Gamma_{av}}{\left(M-1\right)\sin^2 \eta}
%(1+k-\frac{\pi}{4}L_{1/2}^2(-k))}\right\}}{\left(1+\frac{3 N \Gamma_{av}}{\left(M-1\right)\sin^2 \eta} \left(1+k-\frac{\pi}{4}L_{1/2}^2(-k) \right)\right)^\frac{1}{2}} \nn \\
%&& \!\!\!\!\!\!\!
%\approx \text{exp}\left\{\frac{-\frac{3 N^2 \Gamma_{av}\pi }{8\left(M-1\right)\sin^2 \eta}  L_{1/2}^2(-k)}{1+\frac{3 N \Gamma_{av}}{\left(M-1\right)\sin^2 \eta}
%\left(1+k-\frac{\pi}{4}L_{1/2}^2(-k) \right)}\right\}.
%\label{eq74}
%\eeqarr
%\begin{small}
\beqarr
\frac{\exp \left\{\frac{-\frac{3 N^2 \Gamma_{av}\pi }{8\left(M-1\right)\sin^2 \eta}  L_{1/2}^2(-k)}{1+\frac{3 N \Gamma_{av}}{\left(M-1\right)\sin^2 \eta}
\left(1+k-\frac{\pi}{4}L_{1/2}^2(-k) \right)}\right\}}
{\left(1+\frac{3 N \Gamma_{av}}{\left(M-1\right)\sin^2 \eta} \left(1+k-\frac{\pi}{4}L_{1/2}^2(-k) \right)\right)^\frac{1}{2}}
\! \approx \! \exp \! \left\{ \! \frac{-\frac{3 N^2 \Gamma_{av}\pi }{8\left(M-1\right)\sin^2 \eta} L_{1/2}^2(-k)}{1+\frac{3 N \Gamma_{av}}{\left(M-1\right)\sin^2 \eta}
\left(1+k-\frac{\pi}{4}L_{1/2}^2(-k) \right)}\! \right\} \! .
\label{eq74}
\eeqarr
%\end{small}
Using the above results and upon algebraic simplifications, we would obtain the final asymptotic SEP expression as
%Double-column format
%\beqarr
%&& \!\!\!\!\!\!\!\!\!\!\!\!
%P_{s , M\text{-QAM}} \big|_{\Gamma_{av} \ll 1} = 4 \left(\!\!1\!-\frac{1}{\sqrt{M}}\!\!\right)
%Q \left( \sqrt{\frac{3 N^2 \Gamma_{av}\pi L_{1/2}^2 \left(-k\right) }{4}} \right) \nn \\
%&& \!\!\!\!\!\!\!\!\!\!\!\!- 
%\frac {4}{\pi } \left ({1-\frac {1}{\sqrt {M}} }\right)^2
%\left[\frac{\pi}{4}
%-
%\frac{N \pi  L_{1/2}^2 \left(-k\right)  }
%{8 \left(1+ k - \frac{\pi}{4} L_{1/2}^2 \left(-k\right)\right) } \right .\nn \\
%&& \!\!\!\!\!\!\!\!\!\!\!\!
%\left.
%\times 
%\left\{\frac{1}{\left(1+u\right)^{\frac{1}{2}}}
%\left(\tan^{-1}
%\left(\sqrt{1+u}
%\tan \frac{\pi}{4}\right)+
%\frac{\sqrt{1+u}}{2}\tan \frac{\pi}{4}\right)\right\} 
%\right] , \nn \\
%\label{eq75}
%\eeqarr
%\begin{small}
\beqarr
&& \! \! \! \! \! \! \! \! \! \! \! \! \! \! \!
P_{s , M\text{-QAM}} \big|_{\Gamma_{av} \ll 1} =
4 \left(1-\frac{1}{\sqrt{M}}\right)
Q \left( \sqrt{\frac{3 N^2 \Gamma_{av}\pi L_{1/2}^2 \left(-k\right) }{4}} \right)
- \frac {4}{\pi } \left ({1-\frac {1}{\sqrt {M}} }\right)^2
\left[\frac{\pi}{4} \right. \nn \\
&& \! \! \! \! \! \! \! \! \! \!
- \left. \frac{N \pi L_{1/2}^2 \left(-k\right)}
{8 \left(1+k-\frac{\pi}{4} L_{1/2}^2 \left(-k\right)\right)} 
\left\{\frac{1}{\left(1+u\right)^{\frac{1}{2}}}
\left(\tan^{-1}
\left(\sqrt{1+u}
\tan \frac{\pi}{4}\right)+
\frac{\sqrt{1+u}}{2}\tan \frac{\pi}{4}\right)\right\} 
\right] \! \! ,
\label{eq75}
\eeqarr
%\end{small}
where the expression of $u$ is given by 
\beq
u = \frac{M-1}{3 N \Gamma_{av}\left(1+k - \frac{\pi}{4}L_{1/2}^2 \left(-k\right)\right)} \, .
\label{eq76}
\eeq
Furthermore, (\ref{eq61}) and (\ref{eq70}) can be simplified for the cases of $M$-PSK/$M$-QAM constellations for $\Gamma_{av}=0$ to obtain the SEP expression as
\beq
P_{s , M\text{-PSK}} \big|_{\Gamma_{av} = 0} = P_{s , M\text{-QAM}} \big|_{\Gamma_{av} = 0} = M-1/M \, ,
\label{eq77}
\eeq
which is equal to $1/2$ for $M=2$.
%%%%%%%%%%%%%%%%%%%%%%%%%%%%%%%%%%%%%%%%%%%%%%%%%%%%%%%%%%%%%%%%%%%%%%%%%%%%%
%%%%%%%%%%%%%%%%%%%%%%%%%%%%%%%%%%%%%%%%%%%%%%%%%%%%%%%%%%%%%%%%%%%%%%%%%%%%%
%%%%%%%%%%%%%%%%%%%%%%%%%%%%%%%%%%%%%%%%%%%%%%%%%%%%%%%%%%%%%%%%%%%%%%%%%%%%%
%%%%%%%%%%%%%%%%%%%%%%%%%%%%%%%%%%%%%%%%%%%%%%%%%%%%%%%%%%%%%%%%%%%%%%%%%%%%%
\section{Numerical Results}
The performances of the RIS-assisted wireless communication systems under consideration with various system parameters and corroborating the analysis in this paper are presented via numerical results in this section.

\begin{figure}[t]
    \centering
    \includegraphics[height=2.8in,width=4.5in]{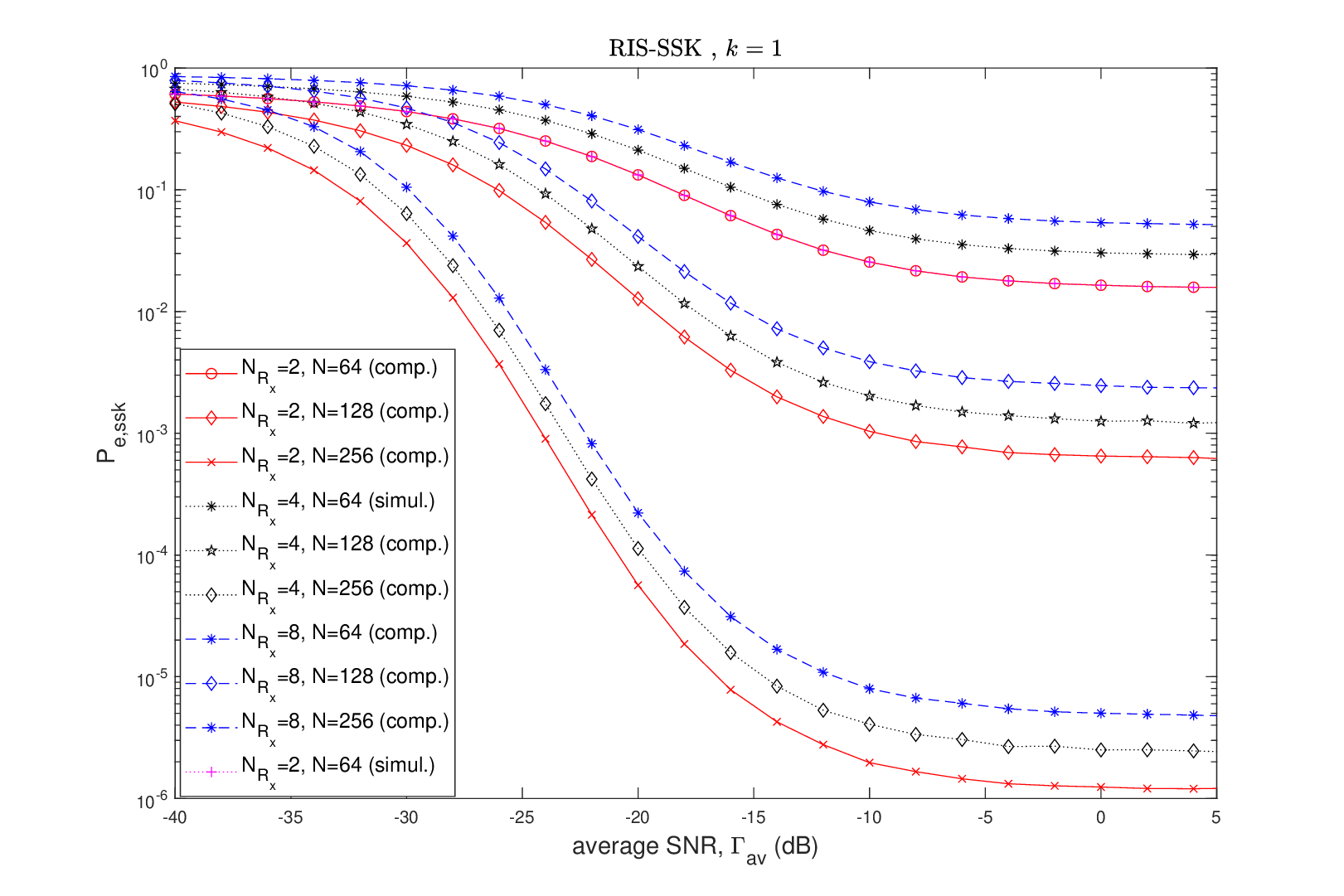}
    \caption{$P_{e,ssk}$ versus $\Gamma_{av}$ for RIS-assisted SSK system for $k=1$, $N_{R_{\text{X}}}=2,4,8$, and $N=64,128,256$.}
    \label{fig1}
\end{figure}
Fig. \ref{fig1} presents the simulation and computation plots of the PED versus the average SNR for the RIS-assisted wireless communication system employing the SSK scheme for varying $N_{R_{\text{X}}}$ and $N$. It is observed that the simulation and computation plots exactly coincide with each other, justifying the correctness of the analytical framework presented in the paper. Furthermore, it is observed that the value of $P_{e,ssk}$ decreases for increasing values of $\Gamma_{av}$, and the variation of the PED is dependent more on $N$ as compared to $N_{R_{\text{X}}}$ which results in a clustering of the plots for a given value of $N$. Additionally, the PED tends to saturate at higher values of $\Gamma_{av}$ as obtained in (\ref{eq29}). It is interesting to observe that the plots of $P_{e,ssk}$ are concave functions of $\Gamma_{av}$ at lower SNR values and are convex functions of $\Gamma_{av}$ at higher SNR values. Thus, there exists a value of $\Gamma_{av}$ which serves as the point of inflection for the plots, and it is observed from Fig. \ref{fig1} that the value of $\Gamma_{av}$ corresponding to the point of inflection reduces with increase in the value of $N$ and decrease in the value of $N_{R_{\text{X}}}$. It is also observed that although the PED values decrease with an increase in the value of $N$, it increases with increasing diversity branches. This can be attributed to the fact that a higher number of antennas at the receiver implies that the target antenna has to be chosen from a set of a higher number of diversity branches which in turn would increase the value of the PED of the system.

\begin{figure}[t]
    \centering
    \includegraphics[height=2.8in,width=4.5in]{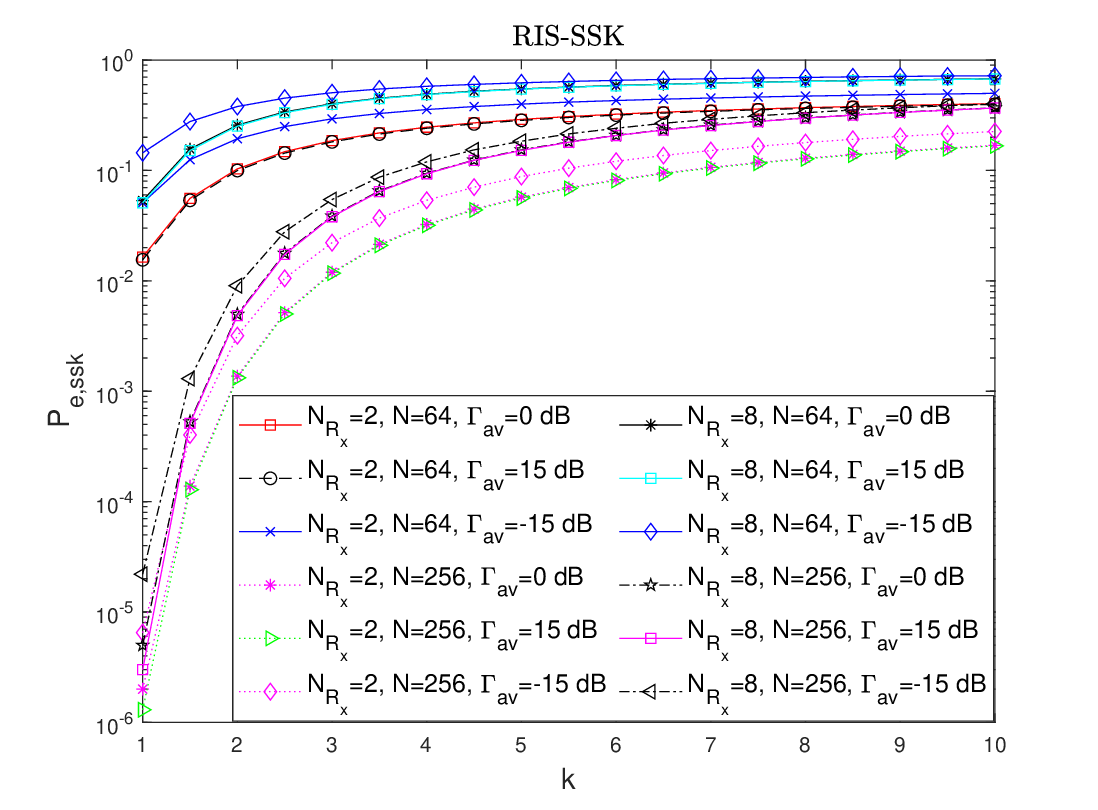}
    \caption{$P_{e,ssk}$ versus Rician parameter $k$ for RIS-assisted SSK system for $N_{R_{\text{X}}}=2,8$, $N=64,256$, and $\Gamma_{av}=-15,0,15$ dB.}
    \label{fig2}
\end{figure}
Fig. \ref{fig2} illustrates the variation of the PED with the Rician factor $k$. It is interesting to observe that the performance of the system in terms of the PED degrades with an increasing Rician factor. This can be attributed to the fact that the LoS component for the non-target antennae becomes prominent along with the target antenna channel, and the effect of the non-target antenna overpowers the performance of the system (similar to the case of the effect of $N$). Moreover, for smaller values of $N$, the variation in the performance of the system is minimal, i.e., the PED values tend to saturate quickly. Additionally, minimal or no performance improvement is associated with using high SNR values at particular $N_{R_{X}}$ and $N $values, and clustering due to the $N$ value is also evident from the plot.  

\begin{figure}[t]
    \centering
    \includegraphics[height=2.8in,width=4.5in]{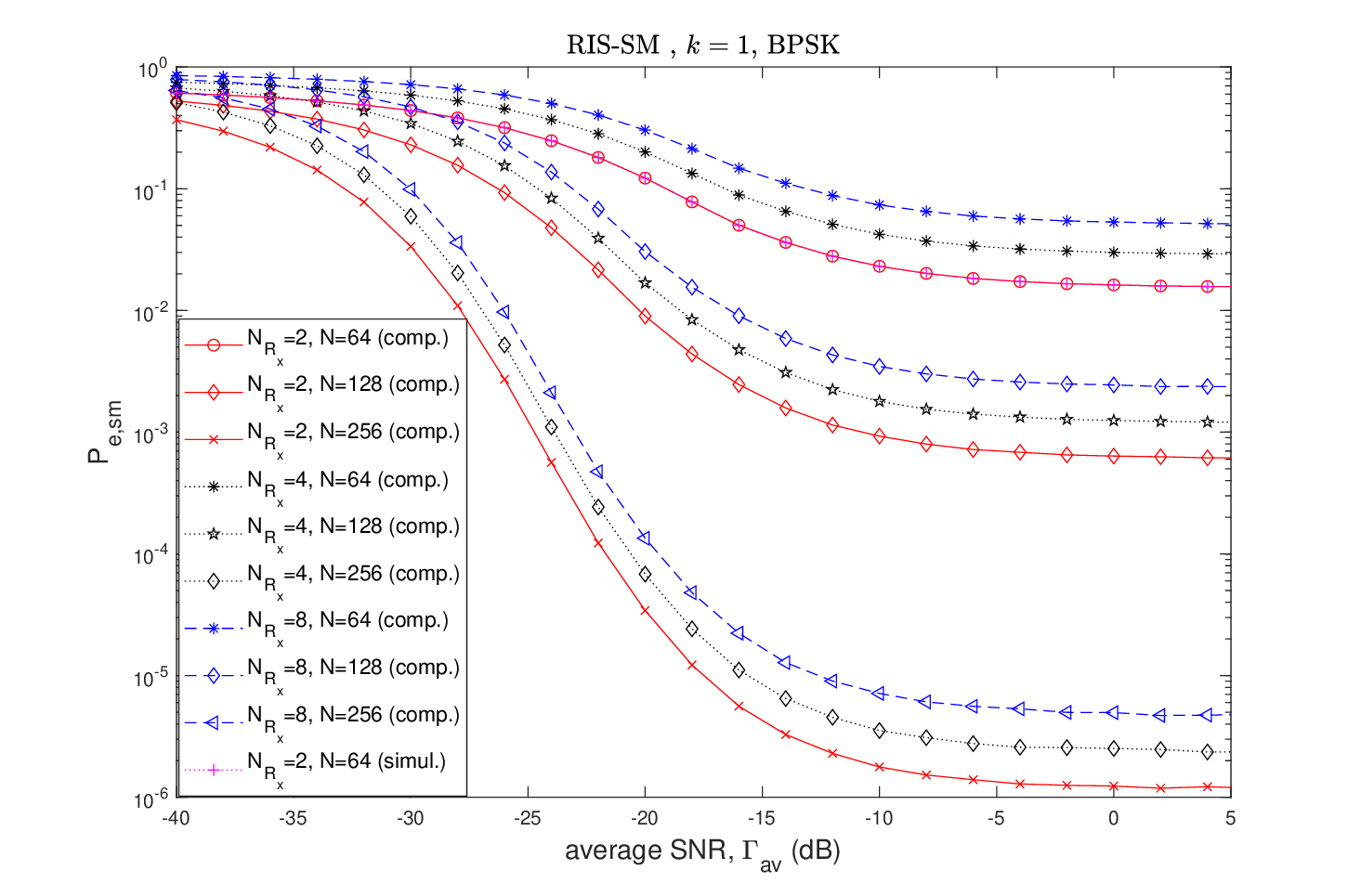}
    \caption{$P_{e,sm}$ versus $\Gamma_{av}$ for RIS-assisted SM system with BPSK modulation for $k=1$, $N_{R_{\text{X}}}=2,4,8$, and $N=64,128,256$.}
    \label{fig3}
\end{figure}
The simulation and computation plots for the PED versus the average SNR for the RIS-assisted SM system with the transmitter employing binary phase shift keying (BPSK) are presented in Fig. \ref{fig3}. The exactness of the computation and simulation plots ensures the correctness of the analysis. Moreover, similar to the case of the RIS-SSK system, $P_{e,sm}$ tends to saturate at higher SNR values with the saturation value obtained in equation (\ref{eq55}). Furthermore, the PED plots are concave functions of $\Gamma_{av}$ at lower values of the average SNR and convex functions at higher values of the average SNR, indicating a point of inflection for the plots. Additionally, increasing the number of the receive diversity antenna has a detrimental effect on the performance, while the exact opposite trend can be observed by increasing the number of reflective surfaces $N$. As compared to the SSK scheme, the clustering effect due to varying values of $N$ is equally evident from the plot.

\begin{figure*}[t]
    \centering
    \includegraphics[height=2.3in,width=6.5in]{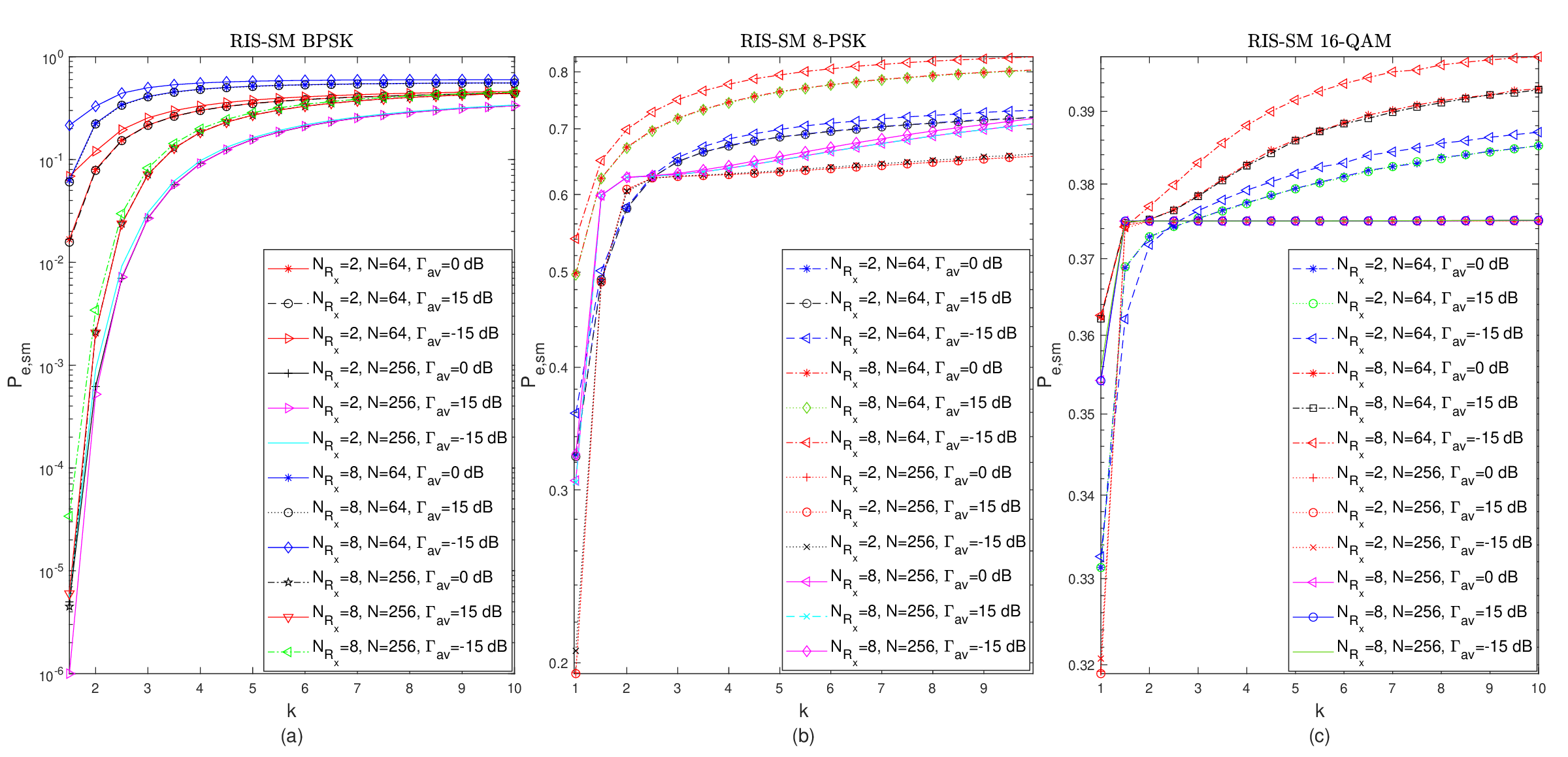}
    \caption{$P_{e,sm}$ versus Rician parameter $k$ for $N_{R_\text{X}}=2,8$, $N=64,256$, and $\Gamma_{av}=-15,0,15$ dB for (a) BPSK; (b) 8-PSK; and (c) 16-QAM constellations for the RIS-assisted SM wireless system.}
    \label{fig4}
\end{figure*}
Fig. \ref{fig4} illustrates the variation of the PED with Rician parameter $k$ for RIS-assisted SM schemes employing BPSK, 8-PSK, and 16-QAM schemes for transmission purposes. As we have seen in the case of SSK modulation, the PED of the system tends to increase with an increasing value of the Rician factor. The degradation of the system performance is substantial for systems with higher values of RIS elements N as compared with those with lower values of N for a given value of $N_{R_{X}}$. Moreover, the system performance degrades by increasing the modulation order of the system, as is evident by comparing the system performance of 8-PSK and 16-QAM with BPSK. Additionally, for a given value of $N_{R_{X}}$ and $N$, there is little or no improvement upon using higher values of $\Gamma_{av}$, as is evident from the plots.

\begin{figure*}[t]
    \centering
	\includegraphics[height=2.4in,width=6.4in]{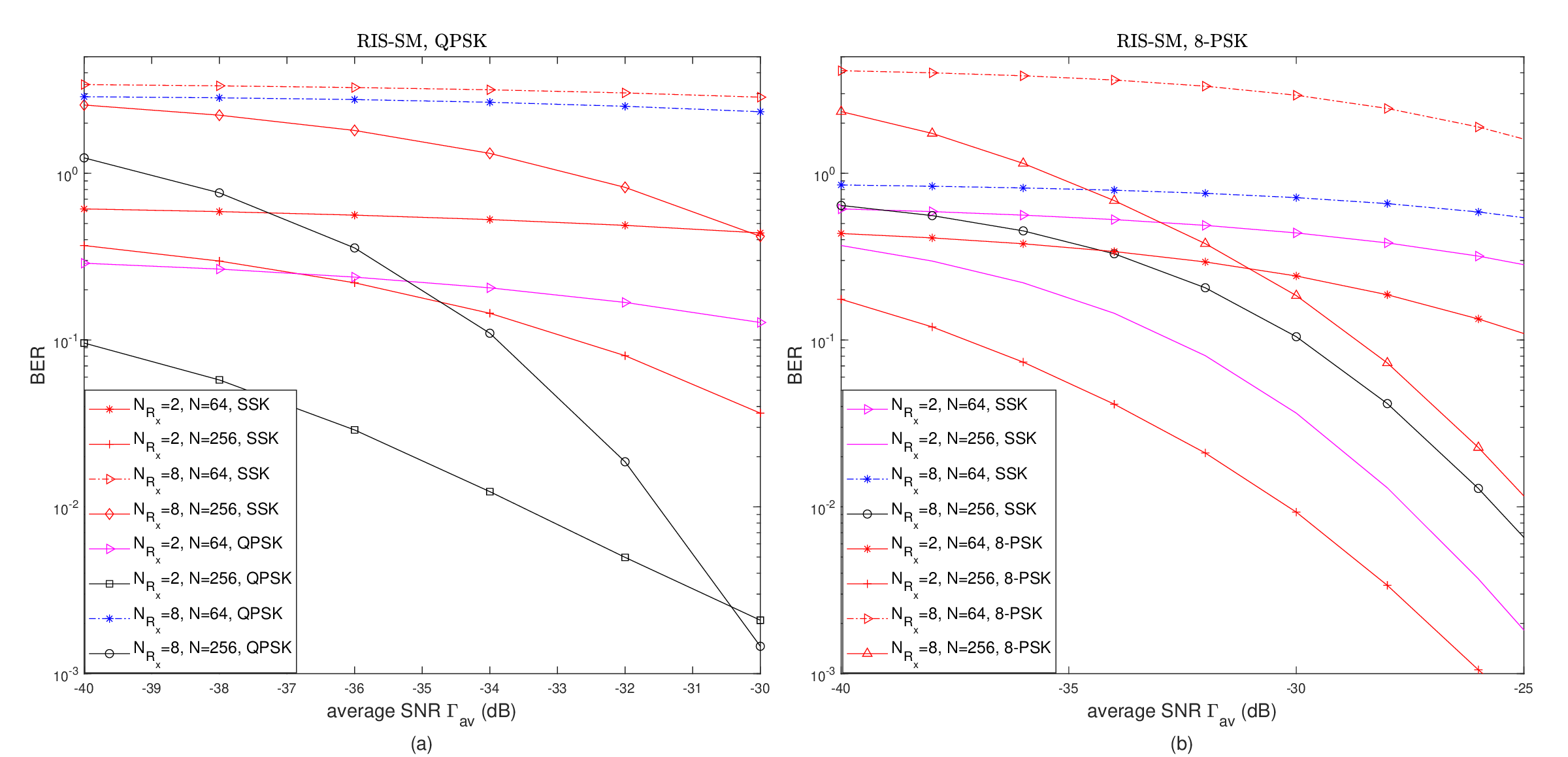}
    \caption{BER versus $\Gamma_{av}$ for RIS-assisted SSK and SM systems with (a) QPSK and (b) 8-PSK modulation schemes for $k=1$, $N_{R_\text{X}}=2,8$, and $N=64,256$.}
    \label{fig7}
    \vspace{-1cm}
\end{figure*}
The plots of the BER versus the average SNR for the RIS-assisted communication system with the SSK scheme and the SM scheme with QPSK and 8-PSK modulations for $m=2$ and varying $N$ and $N_{R_\text{X}}$ are presented in Fig. \ref{fig7}. Additionally, the BER values are obtained using the union-bound technique as given by
\bsub
\beq
P_{b,ssk} \leq N_{R_{\text{X}}} P_{e,ssk}/2 \, ,
\eeq
\beq
P_{b,sm} \approx
\frac{\left( 1 - \left( N_{R_\text{X}}-1 \right) P_{e,sm} \right)
P_{s, M\text{-QAM}/M\text{-PSK}}}
{\log_2 \left( N N_{R_\text{X}} \right)} 
+ \frac{\left( N_{R_\text{X}}-1 \right) P_{e, M\text{-QAM}/M\text{-PSK}}}{2} \, ,
\label{eq:ber_sm}
\eeq
\esub
where the values of $P_{e,ssk}$ and $P_{e,sm}$ are obtained from (\ref{eq29}) and (\ref{eq41}) respectively. It is evident from the plot that the performance of the RIS-assisted SSK modulation scheme is outperformed by both QPSK and 8-PSK modulation schemes. It is noted that the performance of the QPSK scheme is comparatively better than that of the 8-PSK scheme, reaffirming that an increase in the modulation order of the communication system leads to degradation in the performance of the communication system.
%%%%%%%%%%%%%%%%%%%%%%%%%%%%%%%%%%%%%%%%%%%%%%%%%%%%%%%%%%%%%%%%%%%%%%%%%%%%%%%%%%%%%%%%%%%
%%%%%%%%%%%%%%%%%%%%%%%%%%%%%%%%%%%%%%%%%%%%%%%%%%%%%%%%%%%%%%%%%%%%%%%%%%%%%%%%%%%%%%%%%%%
%%%%%%%%%%%%%%%%%%%%%%%%%%%%%%%%%%%%%%%%%%%%%%%%%%%%%%%%%%%%%%%%%%%%%%%%%%%%%%%%%%%%%%%%%%%
%%%%%%%%%%%%%%%%%%%%%%%%%%%%%%%%%%%%%%%%%%%%%%%%%%%%%%%%%%%%%%%%%%%%%%%%%%%%%%%%%%%%%%%%%%%
\section{Conclusions}
In this work, we study a RIS-assisted receive diversity wireless communication system where the transmitter utilizes IM-based schemes for data transmission, and the receiver employs a greedy detection scheme to select the receiver antenna with the maximum received energy for demodulation. The transmitter is considered to be closely situated with respect to the RIS and utilizes SSK or $M$-ary QAM/$M$-PSK-based SM schemes to employ IM. Further, the envelopes of the channel gains between the RIS elements and the receiver are considered to follow i.i.d. Rician distributions statistically. For the employed receiver structure, an analytical framework based on a c.f. approach is presented to derive closed-form expressions for the PED and the SEP of the systems, which serve as the performance metrics of the considered systems. Furthermore, asymptotic expressions of the PED and BER at low and high SNRs reveal the presence of a point of inflection in the PED with respect to the average SNR of the system. The saturation of the PED and union bound of the BER values with respect to the average SNR and the shape parameter of the fading channels are studied via numerical results.
%%%%%%%%%%%%%%%%%%%%%%%%%%%%%%%%%%%%%%%%%%%%%%%%%%%%%%%%%%%%%%%%%%%%%%%%%%%%%%%%%%%%%%%%%%%
%%%%%%%%%%%%%%%%%%%%%%%%%%%%%%%%%%%%%%%%%%%%%%%%%%%%%%%%%%%%%%%%%%%%%%%%%%%%%%%%%%%%%%%%%%%
%%%%%%%%%%%%%%%%%%%%%%%%%%%%%%%%%%%%%%%%%%%%%%%%%%%%%%%%%%%%%%%%%%%%%%%%%%%%%%%%%%%%%%%%%%%
%%%%%%%%%%%%%%%%%%%%%%%%%%%%%%%%%%%%%%%%%%%%%%%%%%%%%%%%%%%%%%%%%%%%%%%%%%%%%%%%%%%%%%%%%%%
\section*{Appendix I\\Derivation of PPED Expression (\ref{eq25})}
The expression of the PPED in (\ref{eq12}) can be expressed as
%Double column format
%\begin{small}
%\beqarr
%\Pr \left\{X_{ssk}<Y_{ssk}\right\} \! \! \! \! &=& \! \! \! \!
%1 - \int_{0}^{\infty} F_{Y_{ssk}} \left(x\right)
%f_{X_{ssk}} \left(x\right) \textnormal{d} x \nn \\
%&=& \! \! \! \! 1 - \int_{0}^{\infty} \int_{0}^{x}
%\frac{\exp \left\{-\frac{y+E_{s}N^2|\mu|^2}{N\sigma_{h}^2E_s + N_0}\right\}}
%{N\sigma_{h}^2 E_s + N_0} \nn \\
%&&   
%\times I_0 \left(\frac{2\sqrt{E_{s}N^2|\mu|^2y}}{N\sigma_{h}^2 E_s + N_0}\right) 
%f_{X_{ssk}}\left(x\right) \textnormal{d}y \textnormal{d}x , \nn \\
%\label{eq17}
%\eeqarr
%\end{small}
%\hspace{-0.15cm}
%\begin{small}
\beqarr
\Pr \left\{X_{ssk}<Y_{ssk}\right\} \! \! \! \! &=& \! \! \! \!
1 - \int_{0}^{\infty} F_{Y_{ssk}} \left(y\right)
f_{X_{ssk}} \left(x\right) \textnormal{d} x \nn \\
&=& \! \! \! \! 1 - \int_{0}^{\infty} \! \! \! \int_{0}^{x}
\frac{\exp \left\{-\frac{y+E_{s}N^2|\mu|^2}{N\sigma_{h}^2E_s + N_0}\right\}}
{N\sigma_{h}^2 E_s + N_0}
I_0 \left(\frac{2\sqrt{E_{s}N^2|\mu|^2y}}{N\sigma_{h}^2 E_s + N_0}\right) 
f_{X_{ssk}}\left(x\right) \textnormal{d}y \textnormal{d}x  ,
\label{eq17}
\eeqarr
%\end{small}
where $F_{X_{ssk}} \left( \cdot \right)$ and $f_{X_{ssk}} \left( \cdot \right)$ denote the cumulative distribution function (c.d.f.) and the probability density function (p.d.f.) of $X_{ssk}$, respectively, and $I_0 (\cdot)$ denotes the modified Bessel function of the zeroth order and the first kind.

Utilizing the series-form expansions of $I_0(x)$ and $\exp \left\{ x \right\}$ as
\beq
I_0(x) = \sum_{\ell=0}^{\infty} \frac{\left( \frac{x^2}{4} \right)^{\ell}}
{\left( \ell ! \right)^2} \quad , \quad
\exp \left\{ x \right\} = \sum_{p=0}^{\infty} \frac{x^p}{p!} \, ,
\label{eq18}
\eeq
in (\ref{eq17}), the expression of $\text{Pr} \left\{X_{ssk}<Y_{ssk}\right\}$ is obtained as
% \beqarr
% && \! \! \! \! \! \! \! \! \! \! \! \!
% \text{Pr} \left\{X_{ssk}<Y_{ssk}\right\} \nn \\
% && \! \! \! \! \! \! \!
% = 1 - \sum_{\ell=0}^{\infty} \sum_{p=0}^{\infty}
% \frac{\left(-1\right)^p \left(E_s N^2 \left|\mu \right|^2 \right)^{\ell}}
% {\left( N \sigma_{h}^2 E_s + N_0 \right)^{2\ell + p + 1}
% \left( \ell !\right)^2 (p!)} \nn \\
% && \!  \times \exp \left\{ -\frac{E_s N^2 \left| \mu \right|^2}
% {N \sigma_{h}^2 E_s + N_0} \right\} 
% \int_{0}^{\infty} \left[ \int_{0}^x y^{\ell+p} \textnormal{d}y \right] f_{X_{ssk}} \left(x\right) \textnormal{d}x \nn \\
% && \! \! \! \! \! \! \!
% = 1 - \sum_{\ell=0}^{\infty} \sum_{p=0}^{\infty}
% \frac{\left(-1\right)^p \left(E_s N^2 \left|\mu \right|^2 \right)^{\ell}}
% {\left( N \sigma_{h}^2 E_s + N_0 \right)^{2\ell + p + 1}
% \left( \ell !\right)^2 p! \left( \ell + p + 1 \right)} \nn \\
% && \quad \times \exp \left\{ -\frac{E_s N^2 \left| \mu \right|^2}
% {N \sigma_{h}^2 E_s + N_0} \right\} 
% \int_{0}^{\infty} x^{\ell + p + 1} f_{X_{ssk}} \left(x\right) \textnormal{d}x. \nn \\
% \label{eq19}
% \eeqarr
%\begin{small}
\beqarr
&& \! \! \! \! \! \! \! \! \! \! \! \!
\text{Pr} \left\{X_{ssk}<Y_{ssk}\right\} \nn \\
&& \! \! \! \! \! \! \!
= 1 - \sum_{\ell=0}^{\infty} \sum_{p=0}^{\infty}
\frac{\left(-1\right)^p \left(E_s N^2 \left|\mu \right|^2 \right)^{\ell}}
{\left( N \sigma_{h}^2 E_s + N_0 \right)^{2\ell + p + 1}
\left( \ell !\right)^2 (p!)}
e^{ -\frac{E_s N^2 \left| \mu \right|^2}
{N \sigma_{h}^2 E_s + N_0}} 
\int_{0}^{\infty} \left[ \int_{0}^x y^{\ell+p} \textnormal{d}y \right] f_{X_{ssk}} \left(x\right) \textnormal{d}x \nn \\
&& \! \! \! \! \! \! \!
= 1 - \sum_{\ell=0}^{\infty} \sum_{p=0}^{\infty}
\frac{\left(-1\right)^p \left(E_s N^2 \left|\mu \right|^2 \right)^{\ell}}
{\left( N \sigma_{h}^2 E_s + N_0 \right)^{2\ell + p + 1}
\left( \ell !\right)^2 p! \left( \ell + p + 1 \right)} 
e^{ -\frac{E_s N^2 \left| \mu \right|^2}
{N \sigma_{h}^2 E_s + N_0}} 
\int_{0}^{\infty} x^{\ell + p + 1} f_{X_{ssk}} \left(x\right) \textnormal{d}x. \nn \\
\label{eq19}
\eeqarr
%\end{small}
It can be observed from (\ref{eq19}) that the integral $\int_{0}^{\infty}x^{\ell +p+1}f_{X_{ssk}}\left(x\right)\textnormal{d}x$ can be computed by using the c.f. of $X_{ssk}$ as
\beqarr
&& \! \! \! \! \! \! \! \! \! \! \! \! \! \! \! \! \! \! \! \! \! \!
\int_{0}^{\infty}x^{\ell+p+1} f_{X_{ssk}}\left(x\right)\textnormal{d}x
= \undb{E} \left[ X_{ssk}^{\ell+p+1} \right]
= \frac{1}{\jmath^{\ell+p+1}} \left.
\left[\frac{\partial^{\ell+p+1}}
{\partial \omega^{\ell+p+1}} \Psi_{X_{ssk}} \left(\jmath \omega \right) \right]\right|_{\omega=0} \, ,
\label{eq20}
\eeqarr
where the expression of $\Psi_{X_{ssk}} \left(\jmath \omega \right)$ is given in (\ref{eq16}).
Moreover, it can be observed from (\ref{eq20}) that the solution of the integral requires the $\ell+p+1$-th differential of the c.f. of $X_{ssk}$. However, from the structure of the c.f. in (\ref{eq16}), a direct differentiation of $\Psi_{X_{ssk}} \left(\jmath \omega \right)$ would be mathematically intractable. Thus, we resolve to employ the Faa di Bruno's formula to solve the differential as
% Double-column format
%\beqarr
%&& \! \! \! \! \! \! \! \! \! \! \! \! \! \! \! \! \! \! \! \!
%\! \!
%\frac{\partial^{\ell+p+1}}{\partial \omega^{\ell+p+1}}
%\Psi_{X_{ssk}} \left(\jmath \omega \right)
%= \left(\ell+p+1 \right)! \, \Psi_{X_{ssk}} \left(\jmath \omega \right) \nn \\
%&& \! \! \! \! \! \!
%\times \hspace{-1cm} \sum_{\begin{array}{c} {\scriptstyle q_1 \ldots, q_{\ell+p+1}} \\
%{\scriptstyle 0 \leq q_1,\ldots,q_{\ell+p+1} \leq \ell+p+1} \\
%{\scriptstyle q_1 + 2q_2 + \ldots + \left( \ell+p+1\right) q_{\ell+p+1} = \ell+p+1}
%\end{array}} \hspace{-1.2cm}
%\prod_{r=1}^{\ell+p+1} \frac{1}{q_r!}
%\left(\frac{H_{ssk}^r \left(\jmath \omega \right)}{r!} \right)^{q_r} \, ,
%\label{eq21}
%\eeqarr
\beqarr
\frac{\partial^{\ell+p+1}}{\partial \omega^{\ell+p+1}}
\Psi_{X_{ssk}} \left(\jmath \omega \right)
= \left(\ell+p+1 \right)! \, \Psi_{X_{ssk}} \left(\jmath \omega \right)
\sum_{S_{q,\ell+p+1}}
% \hspace{-1cm} \sum_{\begin{array}{c} {\scriptstyle q_1 \ldots, q_{\ell+p+1}} \\
% {\scriptstyle 0 \leq q_1,\ldots,q_{\ell+p+1} \leq \ell+p+1} \\
% {\scriptstyle q_1 + 2q_2 + \ldots + \left( \ell+p+1\right) q_{\ell+p+1} = \ell+p+1}
% \end{array}} \hspace{-1.2cm}
\prod_{r=1}^{\ell+p+1} \frac{1}{q_r!}
\left(\frac{H_{ssk}^r \left(\jmath \omega \right)}{r!} \right)^{q_r} \, ,
\label{eq21}
\eeqarr
where the summation over the set $S_{q,\ell+p+1}$ is carried out for all the possible tuples of $q_1,\! \ldots,\!q_{\ell+p+1}$ such that $\sum_{r=1}^{\ell+p+1} rq_r = \ell+p+1$. Moreover, $H_{ssk}^r \left(\jmath \omega \right)$ denotes the $r$-th differential of the function $H_{ssk} \left(\jmath \omega \right)$ defined as
%\begin{small}
\beqarr
H_{ssk} \left(\jmath \omega \right) \dn
\ln \Psi_{X_{ssk}} \left( \jmath \omega \right)
\! \! \! \! &=& \! \! \! \! \frac{\jmath \omega \mu_X^2}{1-2\jmath \omega \left(b+c\right)}
- \frac{1}{2} \ln \left( 1-2\jmath \omega c \right)
% \nn \\
% &-& \! \! \! \!
- \frac{1}{2} \ln \left(1-2\jmath \omega \left(b+c\right) \right) .
\label{eq22}
\eeqarr
%\end{small}
Using (\ref{eq22}), the expression of $H_{ssk}^r \left(\jmath \omega \right)$ can be obtained as
\beqarr
H_{ssk}^r \left(\jmath \omega \right)
\! \! \! \! &=& \! \! \! \!
\frac{\jmath r! \mu_X^2 \left( 2 \jmath \left(b+c \right) \right)^{r-1}}
{\left( 1- 2\jmath \omega \left( b+c \right)\right)^r}
+ \frac{\jmath r! \omega \mu_X^2 \left( 2 \jmath \left(b+c \right) \right)^{r}}
{\left( 1- 2\jmath \omega \left( b+c \right)\right)^{r+1}} \nn \\
&+& \! \! \! \!
\frac{\left(r-1 \right)! \left( 2 \jmath \left(b+c \right) \right)^{r}}
{2 \left( 1- 2\jmath \omega \left( b+c \right)\right)^{r}}
+ \frac{\left(r-1 \right)! \left( 2 \jmath c \right)^{r}}
{2 \left( 1- 2\jmath \omega c \right)^{r}} \, .
\label{eq23}
\eeqarr
Further, from (\ref{eq20}), we need to evaluate the differential at $\omega=0$, implying that
%Double-column format
%\begin{small}
%\beqarr
%H_{ssk}^r (\jmath \omega)\bigg|_{\omega=0} \! \! \! \! \! \! &=& \! \! \! \! 
%\frac{\jmath^{r} \left( r-1 \right)!}{2^{1-r}} \left[r \mu_X^2 \left( b+c \right)^{r-1}
%+ \left( b+c \right)^r+ c^r \right] , \nn \\
%\Psi_{X_{ssk}} (\jmath \omega)\bigg|_{\omega=0} \! \! \! \! \! \! &=& \! \! \! \! 1 \, ,
%\label{eq24}
%\eeqarr
%\end{small}
%\begin{small}
\beq
H_{ssk}^r (\jmath \omega)\bigg|_{\omega=0} =
\frac{\jmath^{r} \left( r-1 \right)!}{2^{1-r}} \left[r \mu_X^2 \left( b+c \right)^{r-1}
+ \left( b+c \right)^r+ c^r \right] \ , \
\Psi_{X_{ssk}} (\jmath \omega)\bigg|_{\omega=0} = 1 \, ,
\label{eq24}
\eeq
%\end{small}
where the variables $\mu_X$, $b$, and $c$ are given in (\ref{eq15}).
Substituting (\ref{eq24}) in (\ref{eq20}) and substituting the result in (\ref{eq16}) we obtain the series-form expression of $\Pr \left\{ X_{ssk} < Y_{ssk}\right\}$ as given in (\ref{eq25}). This completes the proof of Theorem I. \hfill $\blacksquare$
%%%%%%%%%%%%%%%%%%%%%%%%%%%%%%%%%%%%%%%%%%%%%%%%%%%%%%%%%%%%%%%%%%%%%%%%%%%%%%%%%%%%%%%%%%%
%%%%%%%%%%%%%%%%%%%%%%%%%%%%%%%%%%%%%%%%%%%%%%%%%%%%%%%%%%%%%%%%%%%%%%%%%%%%%%%%%%%%%%%%%%%
%%%%%%%%%%%%%%%%%%%%%%%%%%%%%%%%%%%%%%%%%%%%%%%%%%%%%%%%%%%%%%%%%%%%%%%%%%%%%%%%%%%%%%%%%%%
%%%%%%%%%%%%%%%%%%%%%%%%%%%%%%%%%%%%%%%%%%%%%%%%%%%%%%%%%%%%%%%%%%%%%%%%%%%%%%%%%%%%%%%%%%%
\section*{Appendix II\\Derivation of PPED Expression (\ref{eq51})}
Proceeding in similar lines as the RIS-assisted SSK scheme, the expression of the PPED from (\ref{eq37}) is obtained by 
%\begin{small}
\beqarr
\text{Pr} \left\{X_{sm}<Y_{sm}\right\} \! \! \! \! &=& \! \! \! \!
1 - \int_{0}^{\infty} F_{Y_{sm}} \left(y\right)
f_{X_{sm}} \left(x\right) \textnormal{d} x \nn \\
&=& \! \! \! \! 1 - \int_{0}^{\infty} \! \! \! \int_{0}^{x}
\frac{\exp \left\{-\frac{y+E_{s}N^2|\mu|^2|}{N\sigma_{h}^2E_s + N_0}\right\}}
{N\sigma_{h}E_s + N_0}
% \nn \\
% && \hspace{-0.75cm}
% \times
I_0 \left(\frac{2\sqrt{E_{s}N^2|\mu|^2|y}}{N\sigma_{h}^2 E_s + N_0}\right) 
f_{X_{sm}}\left(x\right) \textnormal{d}y \textnormal{d}x .
\label{eq44}
\eeqarr
%\end{small}
Simplifying (\ref{eq44}) by using the series expansion of the modified Bessel function of the zeroth order and the first kind and exponential function as given in (\ref{eq18}), the expression in (\ref{eq44}) can be further simplified as
%Double-column format
%\beqarr
%&& \! \! \! \! \! \! \! \! \! \! \! \!
%\text{Pr} \left\{X_{sm}<Y_{sm}\right\} \nn \\
%&& \! \! \! \! \! \! \!
%= 1 - \sum_{\ell=0}^{\infty} \sum_{p=0}^{\infty}
%\frac{\left(-1\right)^p \left(E_s N^2 \left|\mu \right|^2  \right)^{\ell}}
%{\left( N \sigma_{h}^2 E_s + N_0 \right)^{2\ell + p + 1}
%\left( \ell !\right)^2 (p!)} \nn \\
%&& \hspace{-.6cm}  \times \exp \left\{ -\frac{E_s N^2 \left| \mu \right|^2 }
%{N \sigma_{h}^2 E_s + N_0} \right\} 
%\int_{0}^{\infty} \left[ \int_{0}^x y^{\ell+p} \textnormal{d}y \right] f_{X_{sm}} \left(x\right) \textnormal{d}x \nn \\
%&& \! \! \! \! \! \! \!
%= 1 - \sum_{\ell=0}^{\infty} \sum_{p=0}^{\infty}
%\frac{\left(-1\right)^p \left(E_s N^2 \left|\mu \right|^2\right)^{\ell}}
%{\left( N \sigma_{h}^2 E_s + N_0 \right)^{2\ell + p + 1}
%\left( \ell !\right)^2 p! \left( \ell + p + 1 \right)} \nn \\
%&& \hspace{-.65cm} \times \exp \left\{ -\frac{E_s N^2 \left| \mu \right|^2 }
%{N \sigma_{h}^2 E_s + N_0} \right\} 
%\int_{0}^{\infty} x^{\ell + p + 1} f_{X_{sm}} \left(x\right) \textnormal{d}x. \nn \\.
%\label{eq45}
%\eeqarr
%\begin{small}
\beqarr
\text{Pr} \left\{X_{sm}<Y_{sm}\right\}
\! \! \! \! &=& \! \! \! \! 1 - \sum_{\ell=0}^{\infty} \sum_{p=0}^{\infty}
\frac{\left(-1\right)^p \left(E_s N^2 \left|\mu \right|^2  \right)^{\ell}
e^{-\frac{E_s N^2 \left| \mu \right|^2 }
{N \sigma_{h}^2 E_s + N_0}}}
{\left( N \sigma_{h}^2 E_s + N_0 \right)^{2\ell + p + 1}
\left( \ell !\right)^2 (p!)}
\int_{0}^{\infty} \int_{0}^x y^{\ell+p} \textnormal{d}y f_{X_{sm}} \left(x\right) \textnormal{d}x \nn \\
&=& \! \! \! \!
1 - \sum_{\ell=0}^{\infty} \sum_{p=0}^{\infty}
\frac{\left(-1\right)^p \left(E_s N^2 \left|\mu \right|^2\right)^{\ell}
\exp \left\{ -\frac{E_s N^2 \left| \mu \right|^2 }
{N \sigma_{h}^2 E_s + N_0} \right\} }
{\left( N \sigma_{h}^2 E_s + N_0 \right)^{2\ell + p + 1}
\left( \ell !\right)^2 p! \left( \ell + p + 1 \right)}
\int_{0}^{\infty} \! \! \! x^{\ell + p + 1} f_{X_{sm}} \left(x\right) \textnormal{d}x. \nn \\
\label{eq45}
\eeqarr
%\end{small}
The integral in (\ref{eq45}) can be solved by using the c.f. of $X_{sm}$ as
%Double-column format
%\beqarr
%&& \! \! \! \! \! \! \! \! \! \! \! \! \! \! \! \! \! \! \! \! \! \!
%\int_{0}^{\infty}x^{\ell+p+1} f_{X_{sm}}\left(x\right)\textnormal{d}x
%= \undb{E} \left[ X_{sm}^{\ell+p+1} \right] \nn \\
%&& \qquad \quad
%= \frac{1}{\jmath^{\ell+p+1}} \left.
%\left[\frac{\partial^{\ell+p+1}}
%{\partial \omega^{\ell+p+1}} \Psi_{X_{sm}} \left(\jmath \omega \right) \right]\right|_{\omega=0} \, ,
%\label{eq46}
%\eeqarr
\beq
\int_{0}^{\infty}x^{\ell+p+1} f_{X_{sm}}\left(x\right)\textnormal{d}x
= \undb{E} \left[ X_{sm}^{\ell+p+1} \right]
= \frac{1}{\jmath^{\ell+p+1}} \left.
\left[\frac{\partial^{\ell+p+1}}
{\partial \omega^{\ell+p+1}} \Psi_{X_{sm}} \left(\jmath \omega \right) \right]\right|_{\omega=0} \, ,
\label{eq46}
\eeq
where using Faa di Bruno's formula we have
%Double-column format
%\beqarr
%&& \! \! \! \! \! \! \! \! \! \! \! \! \! \! \! \! \! \! \! \!
%\! \!
%\frac{\partial^{\ell+p+1}}{\partial \omega^{\ell+p+1}}
%\Psi_{X_{sm}} \left(\jmath \omega \right)
%= \left(\ell+p+1 \right)! \, \Psi_{X_{sm}} \left(\jmath \omega \right) \nn \\
%&& \! \! \! \! \! \!
%\times \hspace{-1cm} \sum_{\begin{array}{c} {\scriptstyle q_1 \ldots, q_{\ell+p+1}} \\
%{\scriptstyle 0 \leq q_1,\ldots,q_{\ell+p+1} \leq \ell+p+1} \\
%{\scriptstyle q_1 + 2q_2 + \ldots + \left( \ell+p+1\right) q_{\ell+p+1} = \ell+p+1}
%\end{array}} \hspace{-1.2cm}
%\prod_{r=1}^{\ell+p+1} \frac{1}{q_r!}
%\left(\frac{H_{sm}^r \left(\jmath \omega \right)}{r!} \right)^{q_r} \, ,
%\label{eq47}
%\eeqarr
%\begin{small}
\beq
\frac{\partial^{\ell+p+1}}{\partial \omega^{\ell+p+1}}
\Psi_{X_{sm}} \left(\jmath \omega \right)
= \left(\ell+p+1 \right)! \, \Psi_{X_{sm}} \left(\jmath \omega \right)
\sum_{S_{q,\ell+p+1}}
% \hspace{-1cm} \sum_{\begin{array}{c} {\scriptstyle q_1 \ldots, q_{\ell+p+1}} \vspace{-0.3cm} \\
% {\scriptstyle 0 \leq q_1,\ldots,q_{\ell+p+1} \leq \ell+p+1} \vspace{-0.3cm} \\
% {\scriptstyle q_1 + 2q_2 + \ldots + \left( \ell+p+1\right) q_{\ell+p+1} = \ell+p+1}
% \end{array}} \hspace{-1.2cm}
\prod_{r=1}^{\ell+p+1} \frac{1}{q_r!}
\left(\frac{H_{sm}^r \left(\jmath \omega \right)}{r!} \right)^{q_r} \, .
\label{eq47}
\eeq
%\end{small}
Thus, we need to compute the $r^{th}$ derivative of the function $H_{sm} \left(\jmath \omega \right)$ defined as
%\begin{small}
\beqarr
H_{sm} \left(\jmath \omega \right) \! \! \! \! &\dn& \! \! \! \!
\ln \Psi_{X_{sm}} \left( \jmath \omega \right)
=\frac{\jmath \omega \mu_1^2}{1-2\jmath \omega \left(b_1+c\right)} + \frac{\jmath \omega \mu_2^2}{1-2\jmath \omega \left(b_2+c\right)} \nn \\
&& - \frac{1}{2} \ln \left( 1-2\jmath \omega(b_2 + c)\right) 
- \frac{1}{2} \ln \left(1-2\jmath \omega \left(b_1+c\right) \right) , 
\label{eq48}
\eeqarr
%\end{small}
%Double-column format
%\begin{small}
%\beqarr
%H_{sm} \left(\jmath \omega \right) \dn
%\ln \Psi_{X_{sm}} \left( \jmath \omega \right)&& \!\!\!\!\!\!\!\!\!\!\!=  \!\! \frac{\jmath \omega \mu_1^2}{1-2\jmath \omega \left(b_1+c\right)} + \frac{\jmath \omega \mu_2^2}{1-2\jmath \omega \left(b_2+c\right)} \nn \\
%&& \hspace{-2.9cm} - \frac{1}{2} \ln \left( 1-2\jmath \omega(b_2 + c)\right) 
%- \frac{1}{2} \ln \left(1-2\jmath \omega \left(b_1+c\right) \right) , 
%\label{eq48}
%\eeqarr
%\end{small}
which can be obtained by simple algebraic manipulations as
%\begin{small}
\beqarr
H_{sm}^r \left(\jmath \omega \right)
\! \! \! \! &=& \! \! \! \!
\frac{\jmath r! \omega \mu_1^2 \left( 2 \jmath \left(b_1+c \right) \right)^{r-1}}{\left( 1- 2\jmath \omega \left( b_1+c \right)\right)^{r+1}}  +  \frac{\jmath r! \omega \mu_2^2 \left( 2 \jmath \left(b_2+c \right) \right)^{r-1}}{\left( 1- 2\jmath \omega \left( b_2+c \right)\right)^{r+1}} 
\nn \\  
\hspace{-.8cm} &+& \!\!\!\!\!
\frac{\left(r-1 \right)! \left( 2 \jmath \left(b_1+c \right) \right)^{r}}
{2 \left( 1- 2\jmath \omega \left( b_1+c \right)\right)^{r}}\!
+\! \frac{\left(r-1 \right)! \left( 2 \jmath (b_2 + c)\right)^{r}}
{2 \left( 1- 2\jmath \omega (b_2 + c) \right)^{r}}. 
\label{eq49}
\eeqarr
%\end{small}
Further, to compute the expression of the PPED, we require the values of the derivatives at $\omega=0$ which are obtained as
% Double-column format
%\begin{small}
%\beqarr
%&& \hspace{-2.8cm}
%H_{sm}^r (\jmath \omega)\bigg|_{\omega=0} \! \! \! \! \! \!\!\!\!\!\! = \! \! 
%j^{r} \left( r-1 \right)! 2^{r-1} \left[r \mu_1^2 \left( b_1+c \right)^{r-1} + r \mu_2^2 \left( b_2   +c \right)^{r-1} \right. \nn \\  
%&& + \left. \!\! \left( b_1+c \right)^r+\! (b_2+c)^r \right]  \nn \\
%\Psi_{X_{sm}} (\jmath \omega)\bigg|_{\omega=0} \! \! \! \! \! \! &=& \! \! \! \! 1 \,
%\label{eq50}
%\eeqarr
%\end{small}
\beqarr
H_{sm}^r (\jmath \omega)\bigg|_{\omega=0} \! \! \! \! &=& \! \! \! \! 
j^{r} \left( r-1 \right)! 2^{r-1} \left[r \mu_1^2 \left( b_1+c \right)^{r-1} + r \mu_2^2 \left( b_2   +c \right)^{r-1}
+ \left( b_1+c \right)^r+\! (b_2+c)^r \right] \, , \nn \\
\Psi_{X_{sm}} (\jmath \omega)\bigg|_{\omega=0} \! \! \! \! &=& \! \! \! \! 1 \, .
\label{eq50}
\eeqarr
% \beqarr
% && \! \! \! \! \! \! \! \! \! \! \! \!
% \text{Pr} \left\{X_{sm}<Y_{sm}\right\} \nn \\
% && \! \! \! \! \! \! \!
% = 1 - \sum_{\ell=0}^{\infty} \sum_{p=0}^{\infty}
% \frac{\left(-1\right)^p \left(E_s N^2 \left|\mu \right|^2  \right)^{\ell}}
% {\left( N \sigma_{h}^2 E_s + N_0 \right)^{2\ell + p + 1}
% \left( \ell !\right)^2 (p!)} \nn \\
% && \!  \times \exp \left\{ -\frac{E_s N^2 \left| \mu \right|^2 }
% {N \sigma_{h}^2 E_s + N_0} \right\} 
% \int_{0}^{\infty} \left[ \int_{0}^x y^{\ell+p} \textnormal{d}y \right] f_{X_{sm}} \left(x\right) \textnormal{d}x \nn \\
% && \! \! \! \! \! \! \!
% = 1 - \sum_{\ell=0}^{\infty} \sum_{p=0}^{\infty}
% \frac{\left(-1\right)^p \left(E_s N^2 \left|\mu \right|^2\right)^{\ell}}
% {\left( N \sigma_{h}^2 E_s + N_0 \right)^{2\ell + p + 1}
% \left( \ell !\right)^2 p! \left( \ell + p + 1 \right)} \nn \\
% && \quad \times \exp \left\{ -\frac{E_s N^2 \left| \mu \right|^2 }
% {N \sigma_{h}^2 E_s + N_0} \right\} 
% \int_{0}^{\infty} x^{\ell + p + 1} f_{X_{sm}} \left(x\right) \textnormal{d}x. \nn \\.
% \label{eq16a}
% \eeqarr
Substituting the results in (\ref{eq46})-(\ref{eq50}) followed by algebraic simplifications results in the series-form expression of the PPED of the system to be given in (\ref{eq51}). This completes the proof of Theorem II. \hfill $\blacksquare$
%%%%%%%%%%%%%%%%%%%%%%%%%%%%%%%%%%%%%%%%%%%%%%%%%%%%%%%%%%%%%%%%%%%%%%%%%%%%%%%%%%%%%%%%%%%
%%%%%%%%%%%%%%%%%%%%%%%%%%%%%%%%%%%%%%%%%%%%%%%%%%%%%%%%%%%%%%%%%%%%%%%%%%%%%%%%%%%%%%%%%%%
%%%%%%%%%%%%%%%%%%%%%%%%%%%%%%%%%%%%%%%%%%%%%%%%%%%%%%%%%%%%%%%%%%%%%%%%%%%%%%%%%%%%%%%%%%%
%%%%%%%%%%%%%%%%%%%%%%%%%%%%%%%%%%%%%%%%%%%%%%%%%%%%%%%%%%%%%%%%%%%%%%%%%%%%%%%%%%%%%%%%%%%
\ifCLASSOPTIONcaptionsoff
  \newpage
\fi

\end{document}